\documentclass[%
 reprint,
 amsmath,amssymb,
 aps,
]{revtex4-2}
\usepackage{cleveref}
\usepackage{xcolor,amsmath,amssymb}
\usepackage{bbold}
\usepackage{graphicx}
\usepackage{dcolumn}
\usepackage{bm}

\begin{document}

\preprint{APS/123-QED}

\title{The Spectral Boundary of Block Structured Random Matrices}

\author{Nirbhay Patil}
\email{nirbhay.patil@phys.ens.fr}
 \affiliation{Laboratoire de Physique de l'\'Ecole Normale Supérieure, Paris 75005, France}
\affiliation{Chair of Econophysics and Complex Systems, École polytechnique, 91128 Palaiseau Cedex, France}

\author{Fabián Aguirre-López}%
\affiliation{LadHyX UMR CNRS 7646, École polytechnique, 91128 Palaiseau Cedex, France}
\affiliation{Chair of Econophysics and Complex Systems, École polytechnique, 91128 Palaiseau Cedex, France}

\author{Jean-Philippe Bouchaud}
\affiliation{Capital Fund Management, Paris 75007, France
}%

\affiliation{Chair of Econophysics and Complex Systems, École polytechnique, 91128 Palaiseau Cedex, France}
\affiliation{Academie des Sciences, Paris 75006, France
}%

\date{\today}

\begin{abstract}
Economic and ecological models can be extremely complex, with a large number of agents/species each featuring multiple interacting dynamical quantities. In an attempt to understand the generic stability properties of such systems, we define and study an interesting new matrix ensemble with extensive correlations, generalising the elliptic ensemble. We determine analytically the boundary of its eigenvalue spectrum in the complex plane, as a function of the correlations determined by the model at hand. We solve numerically our equations in several cases of interest, and show that the resulting spectra can take a surprisingly wide variety of shapes. 
\end{abstract}

\maketitle


\section{Introduction}

Ever since Sir Robert May proposed his theory of ecological stability based on the properties of random matrices \cite{may1972will}, there has been an large body of work extending his seminal insight to various ecological situations (see e.g. \cite{macarthur1955fluctuations,mccann2000diversity,may1971stability, allesina2012stability, allesina2015stability, stone2018feasibility, landi2018complexity, baron2020dispersal,baron2022eigenvalue}, and  \cite{akjouj2022complex} for a review) as well as the more recent statistical mechanics approaches to the full non-linear analysis and marginal stability properties of such systems \cite{fyodorov2016nonlinear, biroli2018marginally, altieri2021properties, lorenzana2022well,ros2023generalized}. Interestingly, similar ideas have also appeared in the context of economic networks, see  \cite{moran2019may,liu2020dynamical,dessertaine2022out}, and of random good-exchange networks \cite{nirbhay_rrn} or of more general complex interacting systems \cite{duan2022network}. In the case of trade networks, the structure of the random stability matrix was found to have very specific correlation structure between blocks (see \cite{nirbhay_rrn} and below), for which no existing results could be found in the literature.

In this work we wish to explore the eigenvalue spectra of $NM\times NM$ random matrices $\mathbb{R}$, that can be written as $N\times N$ blocks of size $M\times M$. We assume the elements of these matrices have a non-zero mean along their diagonal block elements,
\begin{align}
    \overline{\mathbb{R}}_{ij}^{\alpha\beta}&= \delta_{ij} \mu_{\alpha \beta}\label{eq:mudef}
\end{align} where the Latin indices $i,j = 1, \dots, N$ indicate which of the $N\times N$ blocks we are in, and the Greek indices $\alpha,\beta=1, \ldots, M$ indicate the element within the block. We further assume that the only non zero correlations are between ${\mathbb{R}}_{ij}^{\alpha\beta}$ and ${\mathbb{R}}_{ij}^{\alpha'\beta'}$ or between ${\mathbb{R}}_{ij}^{\alpha\beta}$ and ${\mathbb{R}}_{ji}^{\alpha'\beta'}$, all other covariances being zero. In other words, the second moments of our matrix ensemble can be expressed as
\begin{align}
    NM\overline{\delta \mathbb{R}_{ij}^{\alpha\beta} \delta\mathbb{R}_{kl}^{\alpha'\beta'}}=\Psi_{\alpha\beta}^{\alpha'\beta'}\delta_{ik}\delta_{jl}+\Upsilon_{\alpha\beta}^{\alpha'\beta'}\delta_{il}\delta_{jk}
    \label{eq:covariances}
\end{align} with the tensors $\Psi$ and $\Upsilon$ having $M^4$ elements each. The aim of this paper is to investigate the spectrum of more general stability matrices, with non trivial block structure induced by the internal state variables. Of particular interest for the stability of the system, we will focus on the boundary of that spectrum in this article, giving access to eigenvalues with extremal real part. 

We show the results of our computations for the boundary of the bulk of the eigenvalue spectra for such matrices in \ref{sec:RMTcalcs} along with a trivial deviation from the elliptical case in \ref{subsec:simplecases}. In \ref{sec:Outliers}, we discuss the possibilities of outlier eigenvalues arising from these systems. In \ref{sec:LVbasic}, we outline the  perturbation analysis for a Lotka-Volterra model, extending it to generic complex systems with minimal constraints in \ref{sec:pertan}. We then show how rotations of the matrix can help include different covariances that have a significant effect on the shape of the bulk of the spectrum in \ref{eq:toyexample}. Then in \ref{sec:decouple} we apply the same transformation on our generic stability matrices to show that our initial findings can apply to these complex systems.  In \ref{sec:2LV} and \ref{sec:RRNexample} we apply our methods to situations known and unknown, inspired from ecological and economical models. Finally, we explore the wide variety of spectra this ensemble can generate in \ref{sec:zoo}  before concluding. 

\section{Methods}
\subsection{Computing the Eigenvalue spectrum for $N \to \infty$ \label{sec:RMTcalcs}}

{We will focus on studying the spectrum of $\mathbb{R}$ in the complex plane, defined as
\begin{equation}
    \rho_N(\omega_x + \textrm{i}\omega_y) = \frac{1}{N} \sum_{i=1}^N \delta(\omega_x - \textrm{Re}[\lambda_i])\delta(\omega_y - \textrm{Im}[\lambda_i]), 
\end{equation}
where $\lambda_i$ corresponds to the $i$-th eigenvalue of matrix $\mathbb{R}$ (possibly complex). In the large $N$ limit, we expect the spectrum to converge to a fixed function, $\rho_N \to \rho$, amenable to be calculated with techniques from statistical physics of disordered systems.}

Let the arrangement of terms in the matrices $\Psi_{\alpha\beta}^{\alpha'\beta'}$ (and $\Upsilon_{\alpha\beta}^{\alpha'\beta'}$) be such that the $(\alpha-1)M+\beta$ row and  $(\alpha'-1)M+\beta'$  column corresponds to the covariance between elements $\mathbb{R}_{ij}^{\alpha\beta}$ and $\mathbb{R}_{ij}^{\alpha'\beta'}$ (or $\mathbb{R}_{ji}^{\alpha'\beta'}$). This gives the joint probability distribution of blocks $\mathbb{R}_{ij}$ and $\mathbb{R}_{ji}$ for any $i$ and $j$ to be
\begin{align}
    P({\bf{R}}_{ij})=\sqrt{\frac{NM}{(2\pi)^{NM}\Delta}}\exp{\left(-\frac{NM}{2}\delta{\bf{R}}_{ij}^T \,{\bf{\Gamma}}^{-1} \, \delta {\bf{R}}_{ij}\right)}
\end{align}
where $\delta {\bf{ R}}_{ij}$  has the terms $ \mathbb{R}_{ij}^{\alpha\beta} - \mu^{\alpha \beta} \delta_{ij}$ arranged in a vector in the sequence $(\alpha-1)M+\beta$, followed by the terms $ \mathbb{R}_{ji}^{\alpha\beta} - \mu^{\alpha \beta} \delta_{ij}$, same as for the matrices $\Psi$ and $\Upsilon$. This ensemble is a generalisation of the one studied in \cite{baron2020dispersal}, which is itself a generalisation of the ensemble introduced in the seminal paper \cite{sommers1988spectrum}. Note that block structured random matrix ensembles have been studied from the mathematical perspective \cite{oraby2007spectral, anderson2008law,aljadeff2015eigenvalues,cicuta2022sparse,cicuta2023sparse} and from the physics perspective \cite{parisi2014soft,cicuta2018unifying,franz2022delocalization,franz2022linear}, since models with $M$ internal degrees of freedom can be useful to understand low energy vibrational modes in glasses.

Let the megamatrix ${\bf {\Gamma}}$ of size $2M^2\times2M^2$ be defined as 
\[
{\bf {\Gamma}}=\begin{pmatrix}
    \Psi & \Upsilon \\ \Upsilon & \Psi
\end{pmatrix},
\] 
and $\Delta=|{\bf {\Gamma}}|=\rm{Det}|\Psi^2-\Upsilon^2|$. The measure of the matrix then becomes $P(\mathbb{R})\approx P({\bf{ R}})^{N^2/2}$ as $\mathbb{R}$ is independent of the block indices $ij$ and all blocks have the same correlation structure. Following the method in \cite{sommers1988spectrum, baron2020dispersal}, we need to calculate the ``electrostatic potential"
\begin{align}
    \Phi(\omega)=-\frac{1}{MN}\langle\ln\rm{Det}|(\mathbb{1}\omega^*-\mathbb{R}^T)(\mathbb{1}\omega-\mathbb{R})|\rangle_J\label{eq:phidef}
\end{align}
which can be represented as a Gaussian Integral as the two matrices multiplied together result in a Hermitian matrix. The identity matrix in \cref{eq:phidef} is of the same size as $\mathbb{R}$, i.e. $MN \times MN$. The resolvent of the matrix is the derivative of this potential
\begin{align}    
G(\omega,\omega^*)=\frac{\partial\Phi(\omega,\omega^*)}{\partial\omega}, \quad 
G^*(\omega,\omega^*)=\frac{\partial\Phi(\omega,\omega^*)}{\partial\omega^*},
\end{align}
and the spectral density is, for any $M$ and $N \to \infty$, the derivative of this resolvent
\begin{align}    
 2\pi\rho={\rm Re}\left[\frac{\partial G}{\partial \omega^*}\right]=
{\rm Re}\left[\frac{\partial G^*}{\partial \omega}\right]
\end{align}
Following the strategy of \cite{baron2020dispersal},
it is then possible to find regions of the complex plane where the eigenvalue density $\rho$ is zero, and regions where the density is surely finite. Solving simultaneously for these regions, we are then able to obtain non-linear equations defining the boundary of the eigenvalue spectrum. Introducing a set of $M^2$ auxiliary variables $q_{\alpha\beta}$ (see Appendix \ref{sec:RMT}), these equations finally read
\begin{align}
 \text{Det}\left[\delta_{\alpha\rho} \delta_{\beta\nu}-\frac{1}{M}\sum_{\gamma,\eta=1}^M q_{\alpha\gamma}\Psi^{\nu \eta}_{\rho\gamma}q^\dagger_{\eta\beta}\right]=0   \label{eq:Kdeteq}
\end{align}
and
\begin{align}
    \begin{aligned}
           \sum_{\alpha'\gamma'\gamma=1}^Mq_{\beta\gamma}q_{\alpha'\gamma'}\Upsilon^{\alpha'\gamma}_{\alpha\gamma'}-M\sum_{\gamma=1}^M(\omega\delta_{\alpha{{\gamma}}}-\mu_{\alpha\gamma} )q_{\beta\gamma}&=-M\delta_{\alpha\beta}.
    \end{aligned}\label{eq:allconds}
\end{align}
This is the central result of the present work.

Note that before taking the determinant we are transforming objects with 4 indices to matrices in 2 indices, by identifying the column number $\rho(M-1)+\nu$ and the row number $\alpha(M-1)+\beta$. 
Based on the specific form of the entries of the correlation matrices, it is now possible to obtain either analytical expressions for the $q$'s in terms of $\omega$ and have an equation relating $\omega_x$ and $\omega_y$ on the complex plane, or numerically solve these equations to get the points on the complex plane that comprise the boundary of the spectrum.

\subsection{Analytical solutions and approximate solutions to simple cases\label{subsec:simplecases}}

\begin{figure}
    \hspace{-0.5cm}\includegraphics[width=1.05\linewidth]{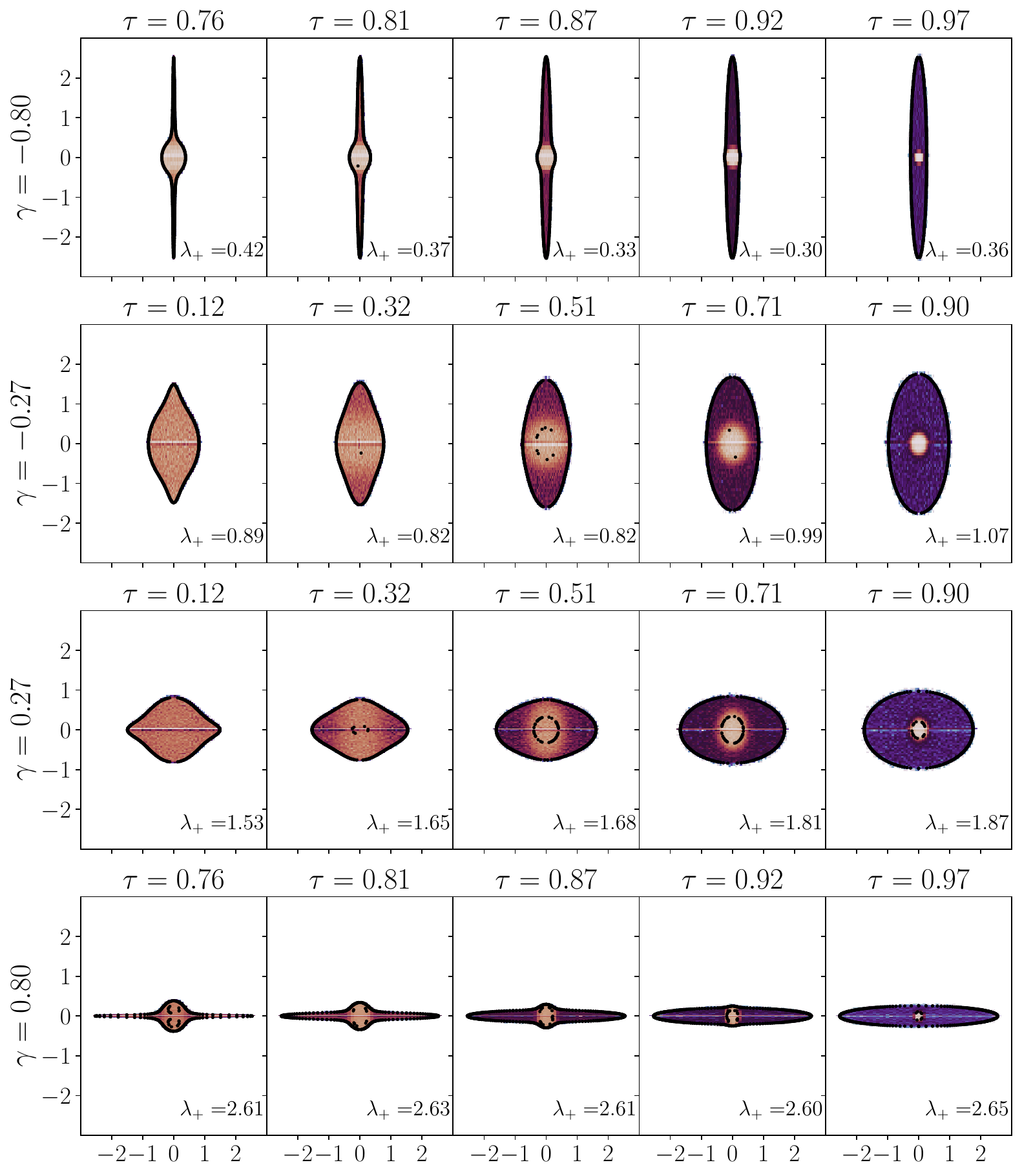}
    
    \caption{Eigenvalue spectra of matrices constructed with constant within-block and inter-block correlations $\sigma^2\tau$ and $\sigma^2\gamma$ respectively, with the variance of each element being $\sigma^2=1$. We see that the leading eigenvalue $\lambda_+$ is a non-monotonic function of the within block correlations $\tau$. The black points along the outside of the spectra are the predictions made by our theoretical results solving for $\omega_x$ and $\omega_y$ using equations \cref{eq:Kdeteq,eq:allconds}. Notice also the inner boundaries that appear to correspond to a change of structure of the corresponding eigenvectors. }\label{fig:spectra_difftauandgam}
\end{figure}

These expressions can in certain conditions simplify to well known cases, for example setting $\mu=0$, we would get a matrix with all elements being i.i.d. random variables  when $\Psi=\mathbb{1}_M$ and $\Upsilon=0$, known to have a circular eigenvalue spectrum \cite{girko1985circular,bordenave2012around}. Instead, setting $\Upsilon=\delta_{\alpha'\beta}\delta_{\alpha\beta'}\tau$ with $-1<\tau<1$ would give an ellipse \cite{sommers1988spectrum}. At $\tau=1$, we would be looking at a real-symmetric Gaussian ensemble, with real eigenvalues having a semi-circular spectrum à la Wigner \cite{wigner1958distribution}. To confirm, let us set $M=1$, $\Psi=\sigma^2$, and $\Upsilon=\sigma^2\tau$. Our general equations simplify into 
\begin{align}
    \begin{aligned}
        \sigma^2 qq^*-1&=0\\
        \sigma^2\tau q^2-q(\omega-\mu)&=-1\\
        \sigma^2\tau (q^*)^2-q^*(\omega^*-\mu)&=-1\\
        \implies \left(\frac{\omega_x-\mu}{1+\tau}\right)^2+\left(\frac{\omega_y}{1-\tau}\right)^2&=\sigma^2
    \end{aligned}
\end{align}
which, as expected, is an ellipse centred at $\mu$ on the real axis. Imagine a slightly more complicated scenario where we have
\begin{align}
    \begin{aligned}
    \Psi_{\alpha\beta}^{\alpha'\beta'}&= \sigma^2\delta_{\alpha\alpha'}\delta_{\beta\beta'}(1-\tau)+\sigma^2\tau, \qquad
    \Upsilon_{\alpha\beta}^{\alpha'\beta'}&=\sigma^2\gamma
\end{aligned}
\end{align}
implying all elements within a block are correlated by the same value $\tau$, and inter-block correlations are also a fixed value $\gamma$. We see the resulting figures that deviate from simple circles and ellipses in \cref{fig:spectra_difftauandgam}. The dependence of the leading eigenvalue on the parameters is non-trivial and non-monotonic, as we see in three of the four rows -- this suggests that for certain disordered systems, higher correlations can lead to higher stability \cite{may1971stability,baron2022eigenvalue}.

\subsection{Outliers\label{sec:Outliers}}

The above analysis only deals with the bulk of the spectrum, while we know that matrices with small non-zero off-diagonal means (aka low-rank perturbations) or wide-spread correlations can have outlier eigenvalues that may also affect the stability of the system being studied \cite{baik2005phase,baron2020dispersal, poley2023eigenvalue}. For the purpose of finding the outliers, we now assume that 
\begin{align}
    \overline{\mathbb{R}}_{ij}^{\alpha\beta}&= \delta_{ij} \mu_{\alpha \beta} + \frac{1}{NM} \mu_{\alpha \beta}^\prime\label{eq:muprimedef}
\end{align}  
giving a matrix $\mu^\prime$ of size $M\times M$ containing the means of the off diagonal elements of each block. Expanding the method given in \cite{baron2020dispersal}, we find that the equations to be solved for the outlier(s) $\omega^\star$ are similar to those for the boundary, namely
\begin{align}    
\begin{aligned}   
 {\rm Det}[\delta_{\alpha\gamma}-\frac{1}{M}\mu^\prime_{\alpha\beta} q_{\beta\gamma}]&=0  \\
\sum_{\alpha'\gamma'\gamma=1}^Mq_{\beta\gamma}q_{\alpha'\gamma'}\Upsilon^{\alpha'\gamma}_{\alpha\gamma'}-M\sum_{\gamma=1}^M(\omega^\star\delta_{\alpha \gamma}-\mu_{\alpha\gamma} )q_{\beta\gamma}&=-M\delta_{\alpha\beta}
\end{aligned}\label{eq:outlierconds}
\end{align}
In addition to these conditions, one needs to check that the solution $\omega^\star$ lies outside the bulk determined by Eqs. \eqref{eq:Kdeteq} and \eqref{eq:allconds}.



\subsection{The Lotka-Volterra case\label{sec:LVbasic}}

In this work we wish to shine some light on the stability matrices of general disordered systems whose linearized dynamics gives rise to a non-trivial correlation structure contained within blocks. Let us motivate our work by starting at a typical Lotka-Volterra equation \cite{lotka1920analytical, biroli2018marginally, altieri2021properties, akjouj2022complex} written for a system of $N$ species as
\begin{align}
    \frac{dx_i}{dt}= x_i \left(\sum_{j=1}^N J_{ij}x_j-x_i+\kappa \right)
\end{align}
where $x_i$ is the abundance of a species $i$ at a particular instance of time, $\kappa$ is a constant signifying the ``carrying capacity'' of the ecosystem for each species (i.e. its fitness), and $J_{ij}$ is the interaction effect of species $j$ on species $i$, assumed to be drawn from a random distribution. A negative interaction signifies that species $j$ is a predator of species $i$ or that they are in competition for similar resources, while a positive one either means $i$ is a predator of $j$ or that they are cooperative specie. 

Performing a linear perturbation around an equilibrium state $\{x_i^*\}_{i=1\dots N}$, i.e. writing $x_i = x_i^* + \epsilon y_i$ with $\epsilon \to 0$, we would get, in a vectorial form:
\begin{align}
    \Dot{\Vec{y}}=\mathbb{L}\Vec{y}\label{eq:matrixeqdef}
\end{align}
where the {\it stability matrix} $\mathbb{L}$ reads
\begin{align}
    \mathbb{L}_{ij}=\delta_{ij}\left(\sum_{j=1}^NJ_{ik}x_k^*-2x_i^*+\kappa\right)+ x_i^*J_{ij}\label{eq:LVmatrix}
\end{align}
Assuming that the state $\{x_i^*\}_{i=1\dots N}$ has a relatively narrow and well-behaved distribution, this becomes a random matrix problem where the stability of \cref{eq:matrixeqdef} depends on the sign of the (real part of the) largest eigenvalues of the matrix $\mathbb{L}$. 

\subsection{Generalised linear perturbation analysis\label{sec:pertan}}

This framework can be extended to more complicated situations in the spirit of constructing generalised models \cite{gross2009generalized}. Consider that instead of having $N$ species with abundances $x_i$, we have $N$ species each having $M$ internal state variables $x_{i}^{(\alpha)}$ with $\alpha=1\dots M$, describing for example sub-species, geographical localization, etc. The time derivatives of each of these variables for a particular species depend in an unspecified manner on all other variables and other species in a pairwise manner, i.e.
\begin{align}
    \begin{aligned}
        \Dot{x}_i^{(\alpha)}=  \sum_{j=1}^N F^{(\alpha)}\left(x_i^{(\{\beta\})},x_j^{(\{\beta\})}; J_{ij}, J_{ji}\right)
    \end{aligned}\label{eq:genLV}
\end{align}
This would boil down to the regular Lotka-Volterra described above when $M=1$, and $F(x_i,x_j)=J_{ij}x_jx_i-{x_i^2}/{N}+{\kappa x_i}/{N}$. 
A linear perturbation of \cref{eq:genLV} around a particular equilibrium state would then lead to
\begin{align}
    \begin{aligned}
        \Dot{y}_i^{(\alpha)}=\sum_{j=1}^N\sum_{\beta=1}^M (\partial_{i,\beta}F^{(\alpha)}_{ij} y_i^{(\beta)}
    +\partial_{j,\beta}F^{(\alpha)}_{ij}y _j^{(\beta)})\end{aligned}\label{eq:generalisedperturbation}
\end{align}
where $y_i^{(\alpha)}$ is the perturbation of variable $x_i^{(\alpha)}$, and \begin{align}
\partial_{k,\beta}F^{(\alpha)}_{ij}:=\frac{\partial F^{(\alpha)}(y_i^{(\{\beta\})},y_j^{(\{\beta\})})}{\partial y_k^{(\beta)}}\Big|_{\Vec{y}^*},
\end{align}
where we dropped the dependence of $F$ on $J_{ij}$'s for readability. Writing \cref{eq:generalisedperturbation} with matrices,
\begin{align}\frac{d}{dt}
    \begin{pmatrix}
        \Vec{y}^{(1)}\\ \Vec{y}^{(2)}\\ \Vec{y}^{(3)}\\.\\.\\.\\ \Vec{y}^{(M)}
    \end{pmatrix}=\begin{pmatrix}
        \mathbb{L}^{(1,1)} & \mathbb{L}^{(1,2)} & \dots & \mathbb{L}^{(1,M)} \\
        \mathbb{L}^{(2,1)} & \mathbb{L}^{(2,2)} & \dots & \mathbb{L}^{(2,M)} \\
        \mathbb{L}^{(3,1)} & \mathbb{L}^{(3,2)} & \dots & \mathbb{L}^{(3,M)} \\
        . & . & \dots & . \\
        .& . & \dots & .\\
        . & . & \dots & .\\
        \mathbb{L}^{(M,1)} & \mathbb{L}^{(M,2)} & \dots & \mathbb{L}^{(M,M)} 
    \end{pmatrix}\begin{pmatrix}
        \Vec{y}^{(1)}\\ \Vec{y}^{(2)}\\ \Vec{y}^{(3)}\\.\\.\\.\\ \Vec{y}^{(M)}
    \end{pmatrix}\label{eq:NxNblocks}
\end{align}
with each $\mathbb{L}^{(\alpha,\beta)}$ being an $N\times N$ matrix given by
\begin{equation}
\mathbb{L}^{(\alpha,\beta)}_{i,j}=\delta_{ij}\sum_{k=1}^N\partial_{i,\beta}F^{(\alpha)}_{ik}+\partial_{j,\beta}F^{(\alpha)}_{ij}.    
\end{equation}
The issue here is that even when the $J_{ij}$ are independent (i.e. $\langle J_{ij}J_{kl}\rangle=\delta_{ik}\delta_{jl}\sigma^2_J$), there are in general non-zero correlations between the blocks $\mathbb{L}^{(\alpha,\beta)}$, 
providing motivation to apply our results in \cref{eq:allconds} to this system. Splitting $\mathbb{L}$ into its average over disorder $\overline{\mathbb{L}}=\langle\mathbb{L}\rangle_J$ plus fluctuations $\delta \mathbb{L}$, we can rewrite the linearized evolution equation as
\begin{align}
    \Dot{{\Vec{y}}}=\left(\overline{\mathbb{L}}+\delta \mathbb{L} \right) \, {\Vec{y}}
\end{align}
where $\Vec{y}$ is the vector of size $MN$ containing all perturbations $\Vec{y}^{(\alpha)}$ from \cref{eq:NxNblocks}, and
\begin{align}
\delta\mathbb{L}^{(\alpha,\beta)}_{i,j}=\delta_{ij}\sum_{k=1}^N g^{(\alpha,\beta)}_{ik} + f^{(\alpha, \beta)}_{ij},
\end{align} 
where we have introduced 
\begin{align}   \begin{aligned}
f^{(\alpha,\beta)}_{ij} &:= \partial_{j,\beta}F^{(\alpha)}_{ij}-\langle\partial_{j,\beta}F^{(\alpha)}_{ij}\rangle_J, \\
g^{(\alpha,\beta)}_{ij} &:= \partial_{i,\beta}F^{(\alpha)}_{ij}-\langle\partial_{i,\beta}F^{(\alpha)}_{ij}\rangle_J
\end{aligned}\label{eq:fandgdefs}
\end{align}
with the overline indicating an average over the couplings $J_{ij}$.

Now, because of the diagonal contribution $\delta_{ij}\sum_{k=1}^N g_{ik}$ to $\delta\mathbb{L}_{i,j}$, one cannot immediately apply the results of section \ref{sec:RMTcalcs} because the covariance structure of $\delta\mathbb{L}$ does not have the same form. We will thus propose an approximation based on the idea of randomly rotating the matrix $\delta\mathbb{L}$ in a way that brings it, to leading order in $N$, to the form considered in section \ref{sec:RMTcalcs}, without changing its eigenvalues that are independent of the chosen basis. However, formally sub-leading correlations between all resulting elements are swept under the rug, which is in general not justified and can change the results even at large $N$. Although framed differently, our approximation here is identical to the one recently proposed in \cite{poley2023eigenvalue}. We will find that for the problems considered later, our approximation is numerically very good. 

\subsection{A Random Rotation Approximation \label{sec:RRA}}

\subsubsection{A toy example\label{eq:toyexample}}

\begin{figure}
    \centering
    \includegraphics[width=\linewidth]{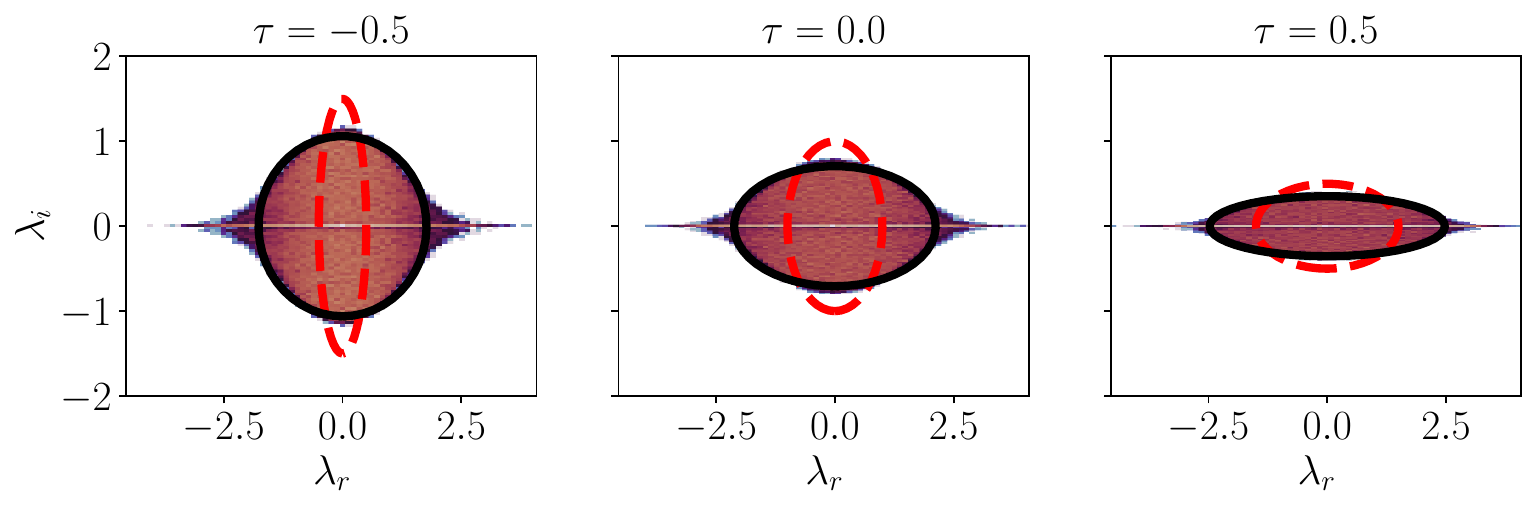}
    \caption{For $N=1000$, $\sigma^2=1/N$, and different values of the correlation between elements $\tau$, we see the actual eigenvalue spectrum of the matrix defined in \cref{eq:ellipseplusdiag}. The dashed red line represents the prediction given by the elliptical law applied on the statistics given the first two terms in \cref{eq:nonrotstats}, and the prediction in black is for the elliptical law applied post rotation on the statistics in \cref{eq:rotationstats}. As we see, the application of the elliptical law to the rotated matrix is a much better approximation of the shape of the spectrum, but still misses some contributions that  significantly deform the shape in the tails along the real axis.}
    \label{fig:rotationexample}
\end{figure}

Let us consider a simpler case with $M=1$ (i.e. no Greek indices) and 
\begin{align}
    \begin{aligned}
        L_{ij}=\delta_{ij}\sum_kf_{ik}+f_{ij},
    \end{aligned}\label{eq:ellipseplusdiag}
\end{align}
with $f_{ij}$ random Gaussian variables with zero mean and 
\begin{align}
    \begin{aligned}
       \overline{f_{ij}f_{kl}}&=\sigma^2\delta_{ik}\delta_{jl}+\tau\sigma^2\delta_{il}\delta_{jk} 
    \end{aligned}
\end{align}
such that
\begin{align}
 \begin{aligned}
             &\overline{L_{ij}L_{kl}}= \sigma^2 \Big( \delta_{ik}\delta_{jl}+\tau \delta_{il}\delta_{jk}  \\
        &+ (\delta_{ij}\delta_{kl}(\delta_{ik}N+\tau)+\delta_{ij}(\delta_{ik}+\delta_{il}\tau)+\delta_{kl}(\delta_{ik}+\delta_{jk}\tau) \Big) 
 \end{aligned}\label{eq:nonrotstats}
\end{align}
Here if we ignore the terms after the first two and only retain the correlations considered in section \ref{sec:RMTcalcs}, we would predict that $L$ has the same spectrum as $f$. As this is wildly inaccurate, we apply a random rotation to split these terms in a way such that they get included into the usual variance and cross-diagonal correlation, and hope that the ignored terms are small enough. Hence we consider the following object:
\begin{align}
    \begin{aligned}
        R_{ab}=\sum_{ij}W_{ai}W_{bj}\left(\delta_{ij}\sum_k f_{ik}+f_{ij} \right)
    \end{aligned}
\end{align}
where $W_{ai}$ is a random rotation matrix in $O(N)$. The corresponding covariances are computed in Appendix \ref{sec:rotappendix} and read, to leading order:
\begin{align}
    \begin{aligned}
        \overline{R_{ab}R_{cd}}&=2\sigma^2\delta_{ac}\delta_{bd}+\sigma^2(1+\tau)\delta_{ad}\delta_{bc}+\mathcal{O}({N}^{-1})
    \end{aligned}\label{eq:rotationstats}
\end{align}
which has the same formal structure as the covariances of $f$ but different coefficients, that reflect the existence of an extra diagonal contribution. The result of such an approximation is shown in \cref{fig:rotationexample} and compared with numerical simulations for 3 different values of $\tau$. We see that although the main features of the spectrum are very well captured, there are visible discrepancies in the tails due to sub-leading, but matrix-wide correlations. 
A better approximation in this case consists in treating the diagonal term in $L_{ij}$ as a random contribution independent of $f_{ij}$, of variance $N \sigma^2$. The boundary of the spectrum of such a model can in fact be computed analytically for all $\tau$ \cite{Pierre}.  

\subsubsection{Block Decoupling\label{sec:decouple}}

Coming back to our general problem, we perform a rotation that takes us from an $M\times M$ matrix whose elements are correlated blocks of size $N\times N$ to an $N\times N$ matrix of uncorrelated blocks of size $M\times M$. For this purpose, consider a random  orthogonal vector basis $W_{(N)}$ of size $N\times N$. 
We thus rotate the matrix $\mathbb{L}$ using the extended vector basis
\begin{align}
    \mathbb{W}=\begin{pmatrix}
        V^{(1,1)} & V^{(2,1)} & V^{(3,1)} & \dots & V^{(N,1)}\\
        V^{(1,2)} & V^{(2,2)} & V^{(3,2)} & \dots & V^{(N,2)}\\
        V^{(1,3)} & V^{(2,3)} & V^{(3,3)} & \dots & V^{(N,3)}\\
        . & . & . & \dots & .\\
        . & . & . & \dots & .\\
        . & . & . & \dots & .\\
        V^{(1,M)} & V^{(2,M)} & V^{(3,M)} & \dots & V^{(N,M)}
    \end{pmatrix}
\end{align}
where $V^{(i,\alpha)}$ represents a $N\times M$ matrix with the column $\alpha$ being equal to the column $i$ of the basis $W$, and all other terms zero, $V^{(i,\alpha)}_{j\beta}=\delta_{\alpha\beta}W_{ij}$. We note that this is just the outer product of two sets of orthogonal basis vectors, $W_{(N)}$ and $\mathbb{1}_{(M)}$. Choosing $\mathbb{1}$ as our second basis can simplify things, but this choice is arbitrary. For some calculations, we use the second basis to also be $W_{(M)}$. Performing the rotation then defines two new matrices
\begin{align}
    \overline{\mathbb{R}}=\mathbb{W}^T\overline{\mathbb{L}}\mathbb{W} \qquad \mathrm{and }\qquad \delta\mathbb{R}=\mathbb{W}^T\delta\mathbb{L}\mathbb{W}.
\end{align}
This gives the constant matrix
\begin{align}
\overline{\mathbb{R}}_{ij}^{\alpha\beta}&=\sum_{\gamma,\delta=1 }^M\sum_{k,l=1}^NV^{(i,\gamma)}_{k\alpha }\overline{\mathbb{L}}_{kl}^{\gamma\delta}V^{(j,\delta)}_{l\beta} = \delta_{ij} \mu_{\alpha \beta}
\end{align} as required, with 
\begin{align}
\mu_{\alpha \beta}&= \sum_{k,l=1}^N W_{ik}W_{il}\left\langle\delta_{kl}\sum_n\partial_{k,\beta}F_{kn}^{(\alpha)}+\partial_{l,\beta}F_{kl}^{(\alpha)}\right\rangle_J 
\label{eq:means}
\end{align}
which gives the mean of the diagonal elements of the block $\mathbb{L}^{(\alpha,\beta)}$ over disorder, whereas the off diagonal elements of the same block would only feature in the outlier calculations. $\overline{\mathbb{R}}$ has blocks of $M\times M$ along the diagonal of an $N\times N$ matrix. Each element of this matrix along the diagonal corresponds to one block of the stability matrix.

The matrix $\delta\mathbb{R}$ is a more complicated object, with
\begin{align}
\delta\mathbb{R}_{ij}^{\alpha\beta}=\sum_{k,l=1}^NW_{i,k}\left(\delta_{kl}\sum_{m=1}^Ng_{km}^{(\alpha,\beta)}+f_{kl}^{(\alpha,\beta)}\right)W_{j,l},
\end{align}
where $f$ and $g$ are as defined in \cref{eq:fandgdefs}. If the functions $F^{(\alpha)}$ are explicitly known, as for the two examples investigated below, it is possible to have the exact expressions for $\Upsilon$ and $\Psi$, so we treat them as $M^2\times M^2$ matrices with possibly known entries. We can therefore regard $\mathbb{R}$ as a random matrix with specific block statistical properties given by \cref{eq:mudef}  and \cref{eq:covariances}. 

Evidently, we require the distributions of the derivatives, $\partial_\beta F^\alpha(J)$, to have well defined first and second moments, and small ($\propto\mathcal{O}({N}^{-1})$) higher order moments for random matrix universality to apply.

\section{Results\label{sec:Apps}}
In this section we apply our theoretical results to two specific cases: the two-island Lotka-Volterra model and a primitive good exchange economy model that two of us have recently introduced \cite{nirbhay_rrn}. As we will show, our approximate random rotation procedure leads to extremely accurate results, suggesting that such an approximation is in fact exact in these cases.

\subsection{Two-island Lotka-Volterra\label{sec:2LV}}

\begin{figure}
    \centering
    \hspace{-0.48cm}\includegraphics[width=1.05\linewidth]{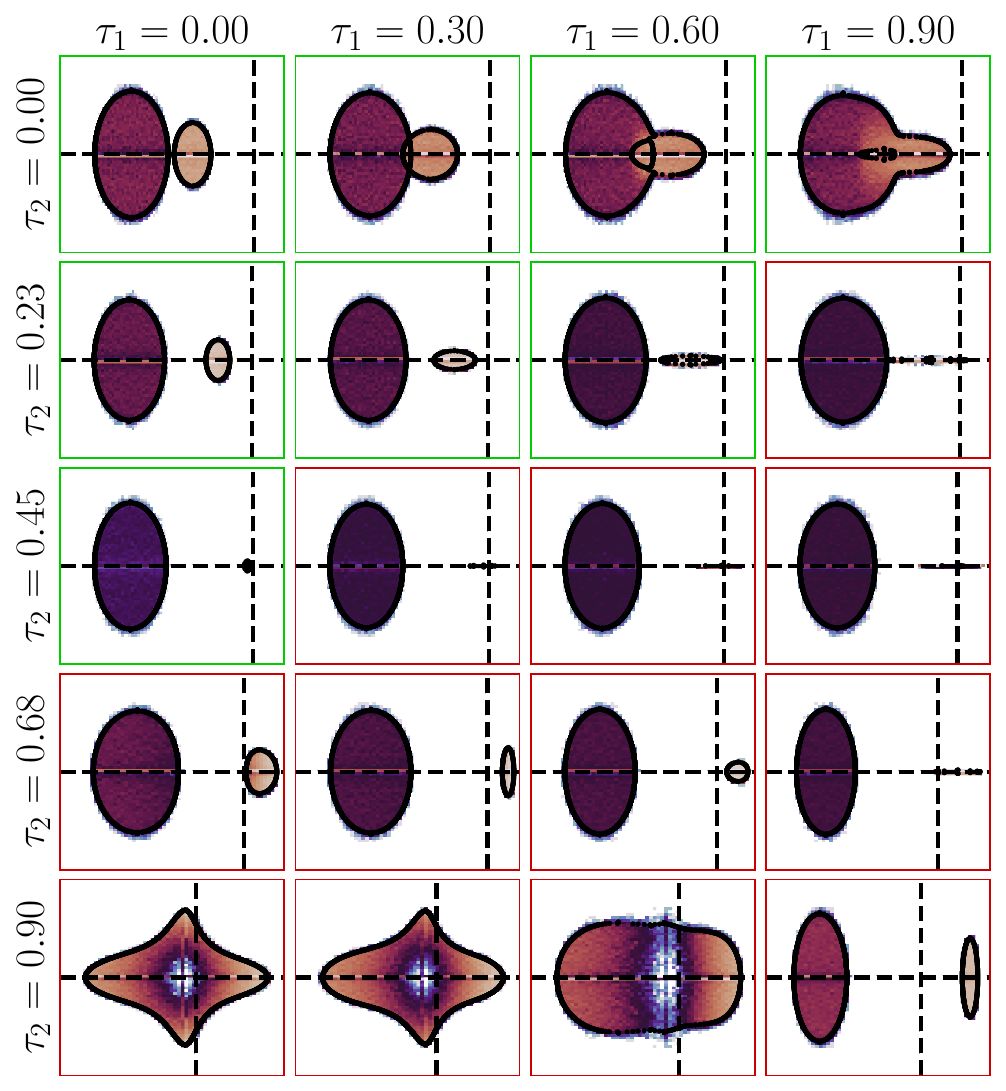}
    \caption{Here we see the spectra of different two island Lotka-Volterra systems, and the boundaries predicted by our theoretical framework given the parameters $\kappa_1=1$, $\kappa_2=2$, and $\overline{J_{ij}J_{kl}}=\delta_{ik}\delta_{jl}/25MN$. The size of the system taken is $N=512$, and the nature of the system dictates of course that $M=2$. We see that in the absence of interaction between the islands the eigenvalue spectrum looks just like two separate circles as expected, while any amount of increase in the saturation term $\tau_2$ or the interaction term $\tau_1$ cause strange changes in the positions and the structures of these shapes. Spectra that feature positive eigenvalues are highlighted in red while the stable cases are in green. }
    \label{fig:2islandLV}
\end{figure}

Consider an ecological system with spatial separation caused by geographical barriers, like species living on two different islands. Using Lotka-Volterra equations, we could write their population dynamics as
\begin{align}
    \begin{aligned}
        \Dot{x}^{(1)}_i&= x_i^{(1)}\left(\sum_{j=1}^NJ_{ij}(x_j^{(1)}+\tau_1 x_j^{(2)})-(x_i^{(1)}+\tau_2 x_i^{(2)})+\kappa_x\right)\\
        \Dot{x}_i^{(2)}&= x_i^{(2)}\left(\sum_{j=1}^NJ_{ij}(x_j^{(2)}+\tau_1x_j^{(1)})-(x_i^{(2)}+\tau_2x_i^{(1)})+\kappa_y\right)
    \end{aligned}
\end{align}
where $x_i^{(1)}$ and $x_i^{(2)}$ are populations of species $i$ on the two different islands respectively. This system gives a non-trivial $4\times4$ correlation matrices. The mean matrix $\mu_{\alpha\beta}$ also features values that are functions of the system parameters $\kappa_x, \kappa_y, \tau_1$, and $\tau_2$. For details on the exact structure and values, check Appendix \ref{appsec:2LV}. We are able to observe the stability of this system semi-analytically by finding the boundary of its spectrum, which matches the spectrum of matrices constructed using the calculated statistics as shown in Fig. \ref{fig:2islandLV}. This model does not exhibit outlier eigenvalues outside of the bulk of its spectrum. 

Notice the presence of ``inner boundaries'' where Eqs. \eqref{eq:allconds} are satisfied but do not correspond to a vanishing of the eigenvalue density. The possibility of such lines is expected on general grounds (see the discussion in Appendix \ref{sec:RMT}). We have found numerically that these correspond, in the two-island case studied here, to a change in the structure of the corresponding eigenvectors, that tend to be localized on one of the island inside the inner boundary and delocalized outside. More work on this aspect would be needed to clarify this observation.

\subsection{A Random Good-Exchange Network\label{sec:RRNexample}}

Consider now the agent-based model proposed in \cite{nirbhay_rrn} as a primitive good-exchange model. Each agent $i=1, \dots, N$ produces $Y=yN$
units of goods for which he/she sets a price $p_i$. Good $j$ has a random utility $J_{ij}$ for agent $i$, who only buys $j$ if $J_{ij}/p_j$ exceeds a certain threshold $S_i$ which is determined by the spending budget of $i$. This spending budget must in turn match the agent's sales. 

More precisely, the matrix $T_{ij}$ dictating whether or not agent $i$ buys from agent $j$ is given by
\begin{align}
    \begin{aligned}
        T_{ij}&=\Theta\left(\frac{J_{ij}}{p_j}-S_i\right)
    \end{aligned}
\end{align}
where $\Theta$ is the Heaviside function (or a smoothed version thereof, see below). The total demand for good $i$ is $D_i = \sum_j T_{ji}$ and the total expenditure of agent $i$ is $E_i = \sum_j p_j T_{ij}$. 

The equations describing the dynamics of $p_i$ and $S_i$ are such that agents increase (decrease) their prices when demand is larger (smaller) than production and adapt their thresholds such as to  balance their books. To wit:
\begin{align}
  \begin{aligned}
     \frac{d\ln p_i}{dt} &= \frac{\kappa}{Y}  (D_i - Y) \\
     \frac{d\ln S_i}{dt} &= \frac{\kappa'}{Y p_i} \left(E_i - p_i D_i\right),
    \end{aligned} \label{eq:cont_rrn}
\end{align}
where $p_i D_i$ is the total sales of agent $i$, i.e. the spending budget. In the following we set $\kappa'=\kappa$. 

As discussed in \cite{nirbhay_rrn}, this system displays a curious phase transition between a stable and an unstable phase, the latter displaying an extremely broad distribution of wealth, which is interesting in this context as a study on inequality. Moreover the system is an  non-trivial extension to generic Lotka-Volterra equations, given its high non-linearity. It also features peculiar terms along the diagonal of each block \cref{eq:ABCDblockdefs} which have a much higher variance than off-diagonal elements. The rotation proposed in sec.\ref{sec:decouple} helps take these variances into consideration with the rest of the terms, giving us a Gaussian approximation to statistics that are in essence of order four. Approximating the step-function as a sharp Fermi-Dirac function and linearizing the equations around a supposed fixed point, we obtain the mean matrix $\mu$ and the $4\times4$ correlation matrices $\Psi$ and $\Upsilon$ that are functions of $\kappa, y, N$, and the distribution of $J_{ij}$ that we assume to be IID exponential random variables. 

\begin{figure}
    \centering
    \includegraphics[width=\linewidth]{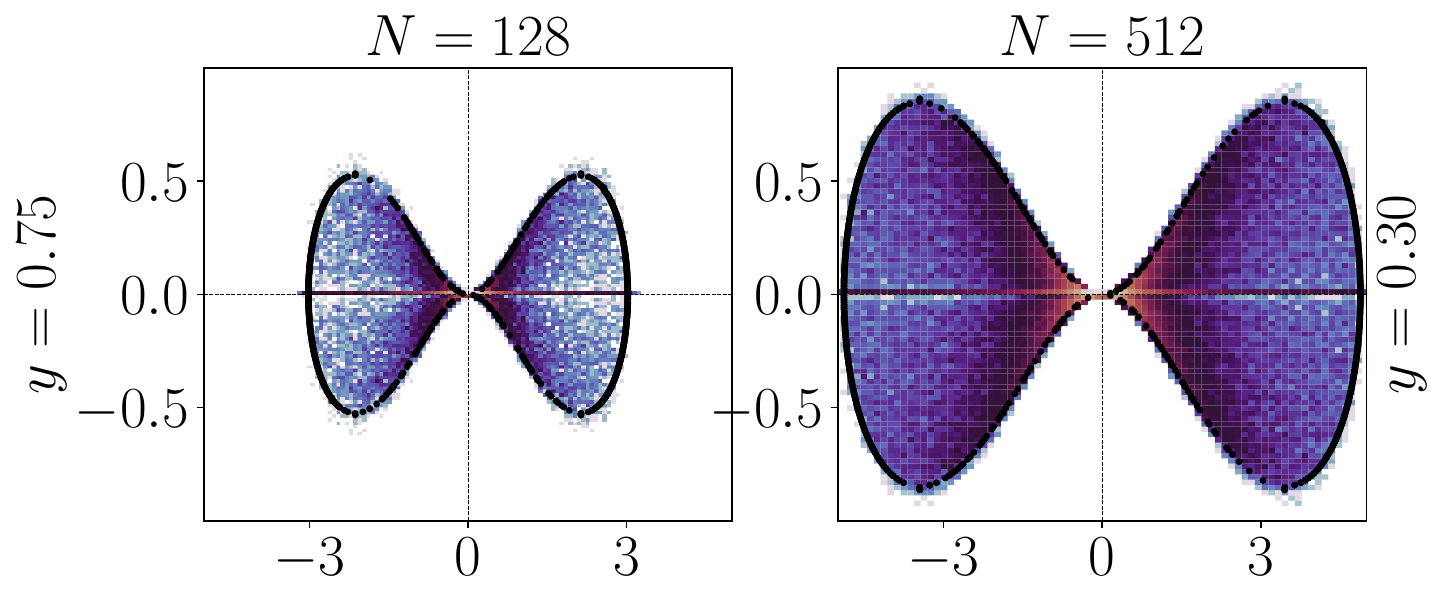}
    \includegraphics[width=\linewidth]{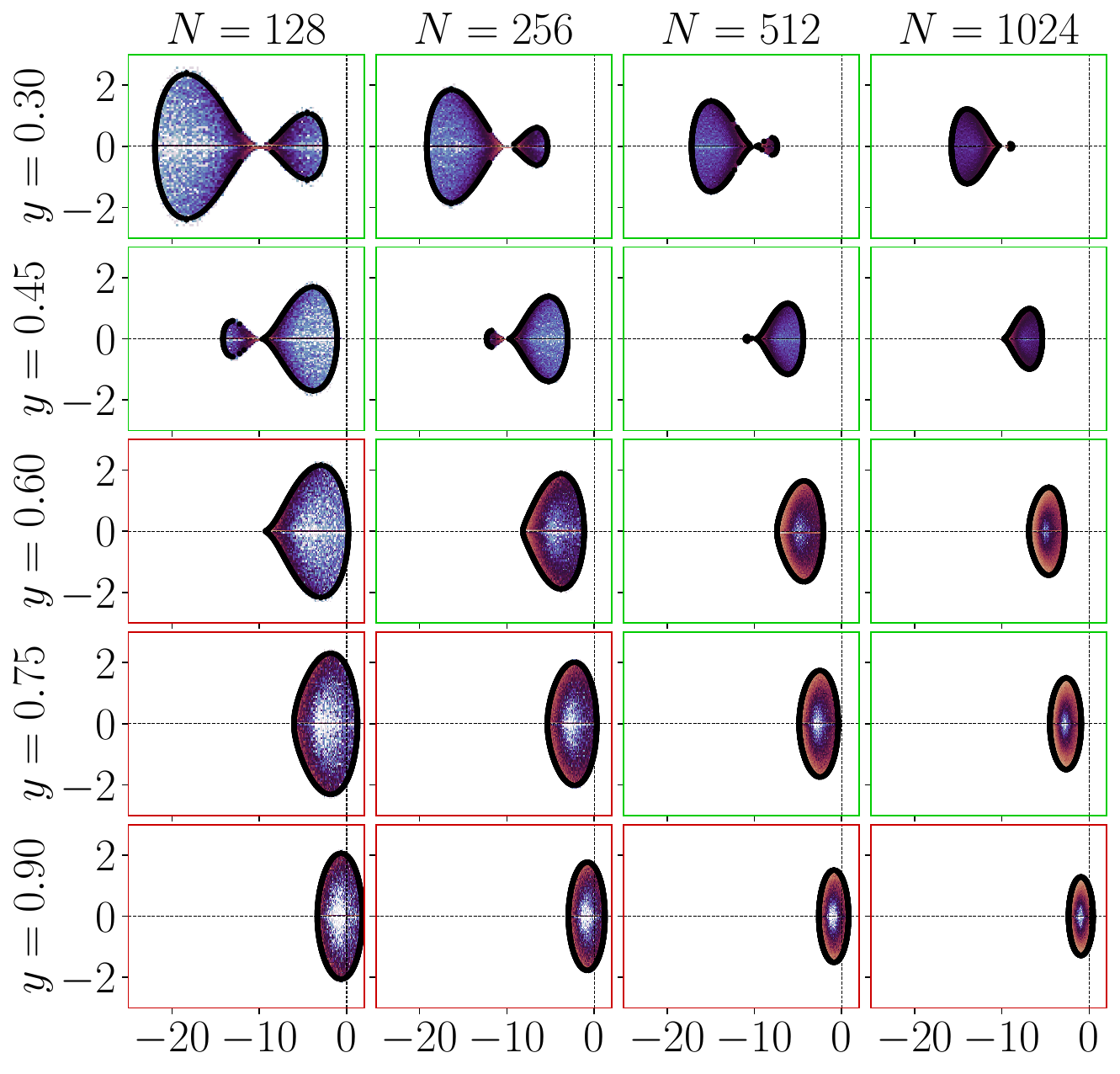}   
    \caption{The eigenvalue spectra of the agent based economic model displays a beautiful butterfly-like shape when the mean $\mu$ is not considered, only scaling the size of the butterfly with the system parameters like $y$ and $N$, while on adding the production dependent mean we see the shape and position of the spectrum change significantly. The system is initially stable, with all eigenvalues sporting negative real parts (plots highlighted in green), but as we increase $y$, the bulk of the spectrum approaches the imaginary axis (plots highlighted in red). This happens faster for smaller systems, which is in line with our initial findings in \cite{nirbhay_rrn}.}
    \label{fig:RRNmodelspectra}
\end{figure}
The resulting matrices $\Psi$ and $\Upsilon$ read to highest order
\begin{align}
\begin{aligned}
    \Psi&= -\begin{pmatrix}
             \begin{pmatrix}
                 2 & 0 \\ -2 & 0
             \end{pmatrix} &
             \begin{pmatrix}
                 0 & 2 \\ 0 & -2
             \end{pmatrix} \\    
             \begin{pmatrix}
                 -2 & 0 \\ 10 & 0
             \end{pmatrix} &
             \begin{pmatrix}
                 0 & -2 \\ 0 & 10\end{pmatrix}\end{pmatrix}\frac{\beta\kappa^2 y\ln y}{3Ny^2} \\
                 \Upsilon&=-\begin{pmatrix}
             \begin{pmatrix}
                 1 & -1 \\ 1 & -1
             \end{pmatrix} &
             \begin{pmatrix}
                 -1 & 1 \\ 3 & -3
             \end{pmatrix} \\    
             \begin{pmatrix}
                 1 & 3 \\ 1 & 3
             \end{pmatrix} &
             \begin{pmatrix}
                 -1 & -3 \\ 3 & 9\end{pmatrix}\end{pmatrix}\frac{\beta\kappa^2 y\ln y}{3Ny^2}
                 \end{aligned}\label{eq:corrs_rrn}
\end{align}
where $\beta = \mathcal{O}(N)$ is a parameter used to define the Heaviside function in a continuous manner as a Fermi-Dirac function, $\theta(x)=({1+e^{-\beta x}})^{-1}$. The mean matrix $\mu$ has the dependence
\begin{align}
    \mu=\frac{\kappa}{2}\begin{pmatrix}
        \ln y-1 & \ln y+1\\ -1-\ln y &3\ln y+1
    \end{pmatrix}\label{eq:mu_rrn}
\end{align}
As shown in \cref{fig:RRNmodelspectra}, we see a spectrum that approaches positive values on the real line as we increase the production capacity $y$, giving an excellent prediction on the transition between egalitarian and oligarchical cases displayed in simulations of the model. The equations \cref{eq:cont_rrn} do not betray any dependence on system size as all terms are of order $N$, only letting us in on the information that $P(J_{ij}>p_jS_i)=y$ at equilibrium. The only thing this piece of information reveals is that we cannot have equilibrium for values of production higher than the number of people ($Y>N$), but we keep seeing this transition occur at values of $Y$ smaller than $N$ in our simulations.  This is visible in \cref{eq:mu_rrn} as well, as the eigenvalue of the constant matrix $\mu$ is $\kappa\ln y$, which approaches 0 as $y$ tends to 1, but should be stable otherwise. We also know there to be one outlier eigenvalue at $\lambda=0$, which corresponds to the invariability of the quantity $p_iS_i$, and does not affect the stability of the system. Through \cref{eq:corrs_rrn} we are finally able to see that the correlations within the stability matrix depend on system size, which manifest in \cref{fig:RRNmodelspectra} to show us how the transition actually happens. Via this method we are able to capture a finite size effect and predict a phase transition in this system as a function of its parameters and its size.

\begin{figure}
    \centering
    \includegraphics[width=\linewidth]{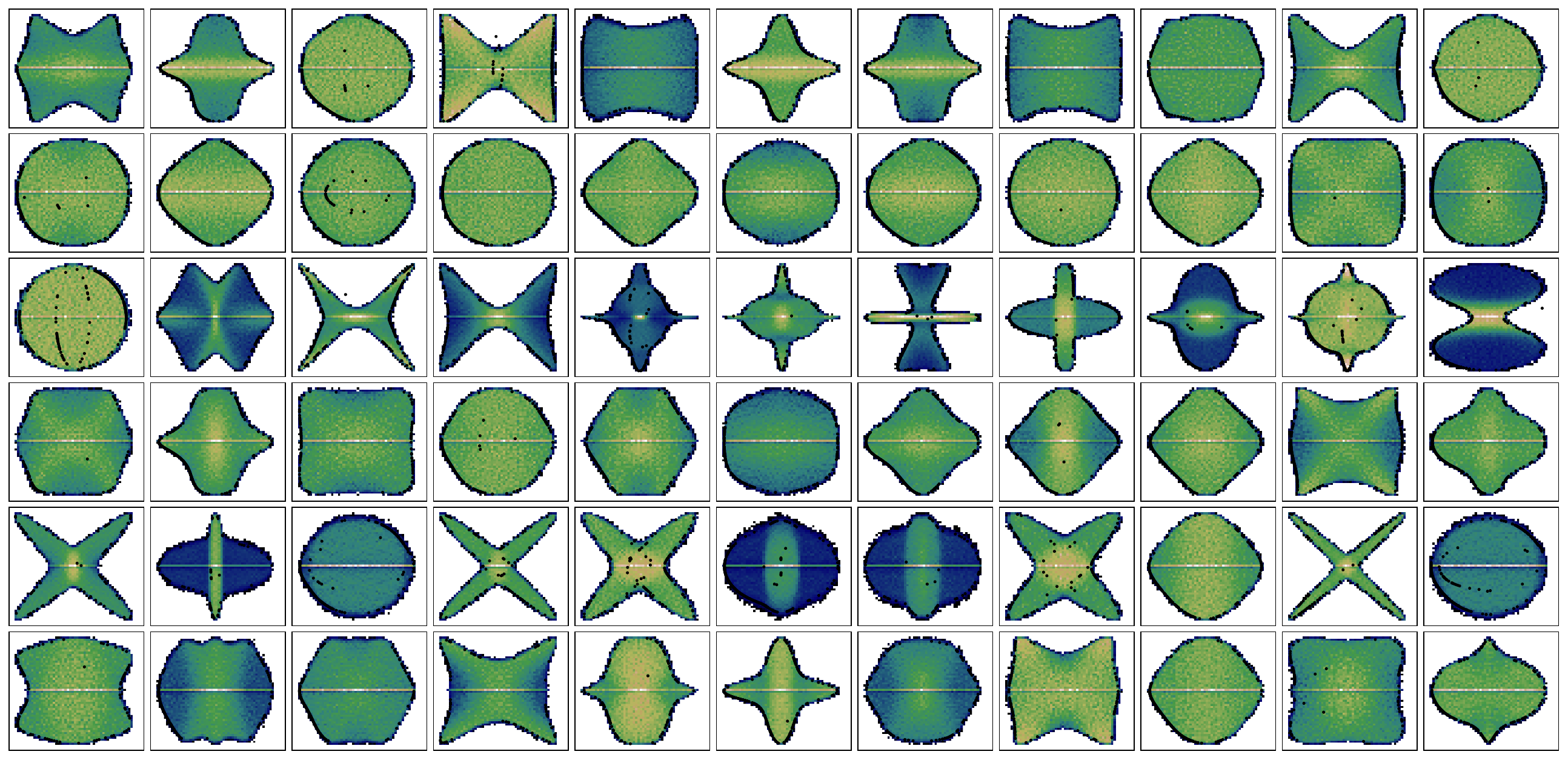}
    \includegraphics[width=\linewidth]{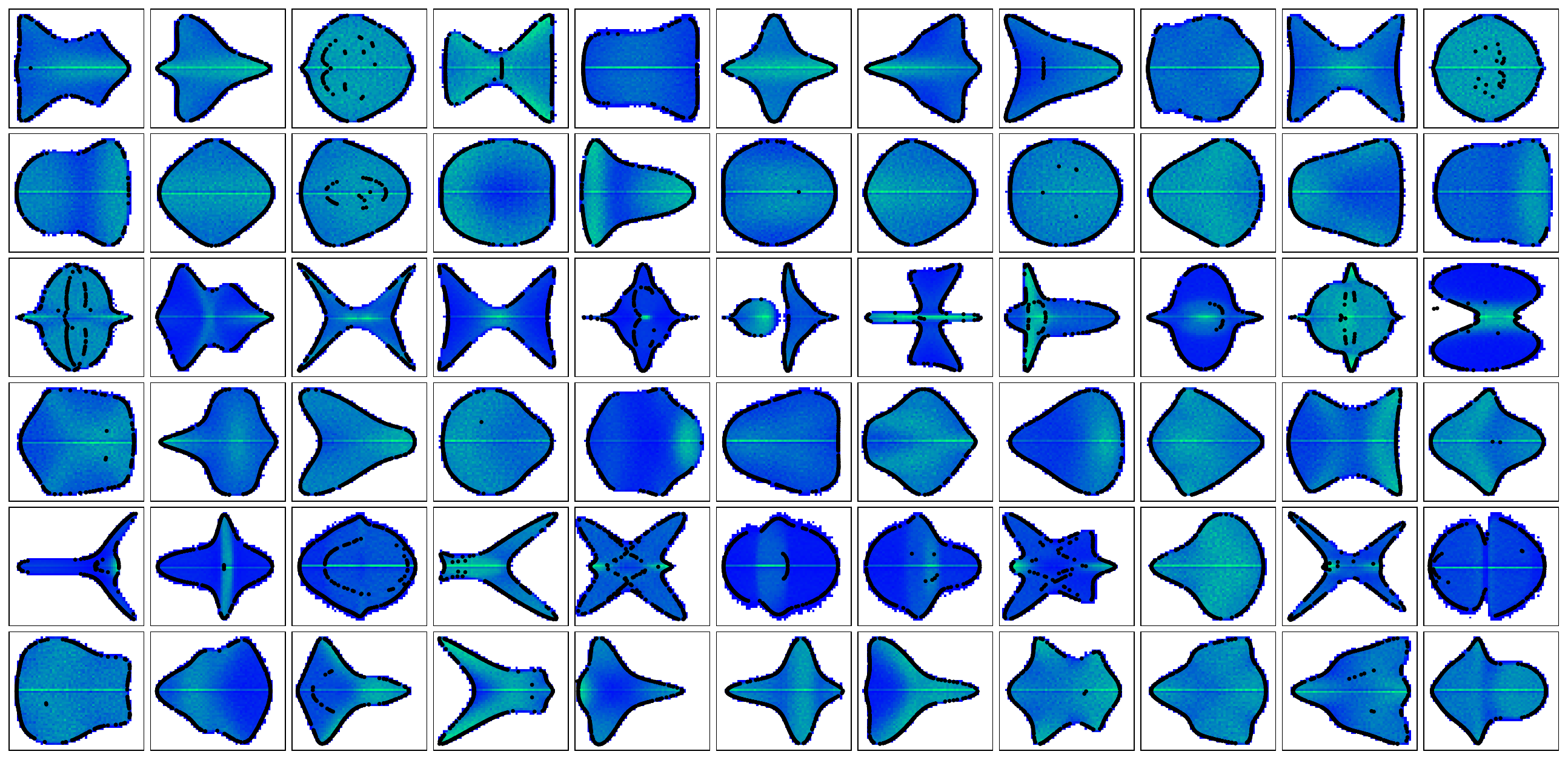}
    \caption{We use randomly generated correlation matrices to explore the space of eigenvalue spectra for $M=2$. The results are visually striking, only limited by the limitations in our process of making the correlation matrices. In the first set of images, the blocks of the matrix have zero means, while the second set has diagonal mean matrices $\mu_{\alpha\beta}$ whose elements are drawn from a Gaussian distribution.}
    \label{fig:random_figs1}
\end{figure}
   
\subsection{A zoo of eigenvalue densities
\label{sec:zoo}}
As a final exploratory exercise to classify the possible shapes of the eigenvalue spectra, we computed the theoretical boundaries of matrices with randomised correlation structures $\Psi$ and $\Upsilon$. To wit, we create such structures by starting off with a diagonal elements with only non-negative values and rotating it in a way that it creates ${\bf{\Gamma}}$ by 
\begin{align}
    \begin{aligned}
       {\bf{\Gamma}}&=\begin{pmatrix}
            \Psi&\Upsilon\\\Upsilon&\Psi
        \end{pmatrix}=\frac{1}{2}\begin{pmatrix}
           O^T&O^T\\O^T&-O^T 
        \end{pmatrix}\begin{pmatrix}
            \Lambda_1&0\\0&\Lambda_2
        \end{pmatrix}\begin{pmatrix}
            O&O\\O&-O
        \end{pmatrix}\\
        &=\frac{1}{2}\begin{pmatrix}
            O^T\Lambda_1O+O^T\Lambda_2O&O^T\Lambda_1O-O^T\Lambda_2O\\O^T\Lambda_1O-O^T\Lambda_2O&O^T\Lambda_1O+O^T\Lambda_2O
        \end{pmatrix}
    \end{aligned}
\end{align}
where $O$ is an $M\times M$ orthogonal vector basis created by starting off with an $M\times M$ random matrix and computing the rotation matrix needed to diagonalize it. This procedure leads to the zoo of shapes shown in \cref{fig:random_figs1}. It would be interesting, but beyond the scope of our paper, to formulate a mathematical theory allowing one to understand the extent of beautiful shapes that can possibly be created this way.

\section{Discussion}

Random matrix theory methods have found extensive utility in theoretical ecology since May's work, and they find themselves right at home in economic models as well. Here we have constructed a framework capable of dealing analytically with a large number of variables, which is often an unavoidable feature of agent-based models. As we deal with larger and larger structures of matrices, it becomes difficult to point out which correlations lead to what sort of change in shape and whether they are directly responsible for a change in stability. We are however able to construct the shapes the spectra would exhibit with this knowledge, making stability analyses possible for a large variety of models. There are other methods to calculate the spectral density of block structured Lotka-Volterra-like systems, like using the cavity method as in \cite{grilli2016modularity}. In our analysis, we hope to have been able to solve such systems for a wider diversity of correlation values and for any number of variables.

It is possible to extend our work in various ways, for instance by adding fine structure to the correlations as defined in \cite{poley2023eigenvalue}, or working with models where the number of variables is not fixed to $N$ as above, but is different for each variable, as it could be if we were dealing with multiple kinds of agents. This could arise in an economic system comprised of $N_1$ firms interacting with $N_2$ banks and $N_3$ people. Such a system would have $M=3$, but the blocks in the stability matrix would all be of different sizes, requiring a more elaborate rotation to arrive at the kind of structure we now know the results for.

Finally, let us mention two open problems that we have not been investigated in details in the present work. One concerns the appearance of ``inner boundaries" within the domain of existence of eigenvalues, which seems to be associated with a change of the structure of the corresponding eigenvectors. How this can be understood analytically has eschewed us at this point. Second is the study of the large $M$ limit of Eqs. \eqref{eq:allconds}, which should allow for a deeper understanding of the analytical properties of the spectra generated by our family of models -- which, as illustrated in Fig. \ref{fig:random_figs1}, can take amazingly different forms and shapes.  

\section*{Acknowledgements}

We would like to thank Y. Fyodorov, P. Bousseyroux, J. Baron and G. Biroli for fruitful discussions on the subject matter. The authors would also like to thank G. Lorenzana, S. Elomari, and J. Garnier-Brun for their input and suggestions. This research was conducted within the Econophysics \& Complex Systems Research Chair, under the aegis of the Fondation du Risque, the Fondation de l’\'Ecole polytechnique, the \'Ecole polytechnique and Capital Fund Management.
 
\nocite{*}

            \begin{widetext}
\appendix 
\section{Spectrum computations}\label{sec:RMT}
\subsection{The electrostatic potential and saddle point integration}
Let the arrangement of terms in the matrices $\Psi_{\alpha\beta\alpha'\beta'}$ and $\Upsilon_{\alpha\beta\alpha'\beta'}$ be such that the $(\alpha-1)M+\beta$ row and  $(\alpha'-1)M+\beta'$ column corresponds to the covariance between elements $\mathbb{R}_{ij}^{\alpha\beta}$ and $\mathbb{R}_{ij}^{\alpha'\beta'}$. This gives the joint probability distribution of blocks $\mathbb{R}_{ij}$ and $\mathbb{R}_{ji}$ for any $i$ and $j$ to be
\begin{align}
    P(\Vec{ R_{ij}})=\sqrt{\frac{NM}{(2\pi)^{NM}\rm{Det}|\Psi^2-\Upsilon^2|}}\exp{\left(-\frac{NM}{2}(\Vec{R_{ij}}^T-\Vec{\mu}^T)\begin{pmatrix}
        \Psi & \Upsilon \\ \Upsilon & \Psi
    \end{pmatrix}^{-1}(\Vec{R_{ij}}-\Vec{\mu})\right)}
\end{align}
where $\Vec{ R_{ij}}$  has the terms $ \mathbb{R}_{ij}^{\alpha\beta}$ arranged in a vector in the sequence $(\alpha-1)M+\beta$, followed by the terms $ \mathbb{R}_{ji}^{\alpha\beta}$,  same as the matrices $\overline{\sigma}^2$ and $\overline{\gamma}^2$. The measure of the matrix then becomes $P(\mathbb{R})\approx P(\Vec{ R})^{N^2/2}$ as $\mathbb{R}$ is independent in block indices $ij$ and all blocks have the same correlation structure. Let the megamatrix ${\bf {\Gamma}}$ be defined as ${\bf {\Gamma}}=\begin{pmatrix}
    \Psi & \Upsilon \\ \Upsilon & \Psi
    
\end{pmatrix}$, and $\Delta=|{\bf {\Gamma}}|=\rm{Det}|\Psi^2-\Upsilon^2|$. The average over disorder of any quantity $X$ would be
\begin{align}
\begin{aligned}
    \langle X\rangle_J&=\int\left(\prod_{i<j,\alpha\beta\alpha'\beta'}d\mathbb{R}_{ij}^{\alpha\beta}d\mathbb{R}_{ji}^{\alpha'\beta'}\sqrt{\frac{NM}{(2\pi)^{MN}\Delta}}\right)X\exp\left[-\sum_{i<j}\frac{NM}{2}(\Vec{ R}_{i,j}^T-\Vec{\mu}^T){\bf {\Gamma}}^{-1}(\Vec{ R}_{i,j}-\Vec{\mu})\right]\\
    \implies \langle X\rangle_J&\approx\int\left(\prod_{ij\alpha\beta\alpha'\beta'}d\mathbb{R}_{ij}^{\alpha\beta}d\mathbb{R}_{ji}^{\alpha'\beta'}\sqrt{\frac{NM}{(2\pi)^{NM}\Delta}}\right)X\exp\left[-\sum_{ij}\frac{NM}{4}(\Vec{ R}_{i,j}^T-\Vec{\mu}^T){\bf {\Gamma}}^{-1}(\Vec{R}_{i,j}-\Vec{\mu})\right]
\end{aligned}
\end{align}
Following the method in \cite{sommers1988spectrum}, we need to calculate the electrostatic potential

\begin{align}
    \Phi(\omega)=-\frac{1}{NM}\langle\ln\rm{Det}|(\mathbb{I}\omega^*-\mathbb{R}^T)(\mathbb{I}\omega-\mathbb{R})|\rangle_J
\end{align}
which can be represented as a Gaussian Integral as the two matrices multiplied together result in a Hermitian matrix
\begin{align}
    \Phi(\omega)=-\frac{1}{NM}\ln\left\langle\int\left[ \prod_{i\alpha}\frac{d^2z^\alpha_i}{\pi}\right]\exp\left\{-\epsilon\sum_{i\alpha}|z_i^\alpha|^2-\sum_{ijk\alpha\beta\gamma}z_i^{\alpha*}(\omega^*\delta_{ik}\delta_{\alpha\gamma}-(\mathbb{R}^T)^{\alpha\gamma}_{ik})(\omega\delta_{kj}\delta_{\gamma\beta}-\mathbb{R}^{\gamma\beta}_{kj})z^\beta_j\right\}\right\rangle_J\label{eq:Phigaussexp}
\end{align}
where $\epsilon$ is an infinitesimally small positive value added to $\omega$ to avoid facing ill defined quantities arising when $\omega$ is an eigenvalue of $\mathbb{R}$. We are also relying heavily on \cite{sommers1988spectrum}'s proof via replicas that the annealed and quenched averages are the same in the large $N$ limit. Performing the complex equivalent of an Hubbard-Stratonovich transformation on \cref{eq:Phigaussexp}, we get
\begin{align}
    \begin{aligned}
\exp(-MN\Phi(\omega))=\Bigg\langle\int\prod_{i\alpha}\left(\frac{d^2z_i^\alpha d^2y_i^\alpha}{2\pi^2}\right)&\exp\left[-\sum_{i\alpha}(y_i^{\alpha*}y_i^{\alpha}+\epsilon z_i^{\alpha*}z_i^{\alpha})\right]\exp\left[i\sum_{ij\alpha\beta}z_i^{\alpha*}(\mathbb{R}_{ji}^{\beta\alpha}-\omega^*\delta_{ij}\delta_{\alpha\beta})y_j^{\beta}\right]\\
&\times \exp\left[i\sum_{ij\alpha\beta}y_i^{\alpha*}(\mathbb{R}_{ij}^{\alpha\beta}-\omega\delta_{ij}\delta_{\alpha\beta})z_j^{\beta}\right]\Bigg\rangle_J\\
=\Bigg\langle\int\prod_{i\alpha}\left(\frac{d^2z_i^\alpha d^2y_i^\alpha}{2\pi^2}\right)&\exp\left[-\sum_{i\alpha}(y_i^{\alpha*}y_i^{\alpha}+\epsilon z_i^{\alpha*}z_i^{\alpha})\right]\exp\left[i\sum_{ij\alpha\beta}z_j^{\beta*}(\mathbb{R}_{ij}^{\alpha\beta}-\omega^*\delta_{ij}\delta_{\alpha\beta})y_i^{\alpha}\right]\\
&\times \exp\left[i\sum_{ij\alpha\beta}y_i^{\alpha*}(\mathbb{R}_{ij}^{\alpha\beta}-\omega\delta_{ij}\delta_{\alpha\beta})z_j^{\beta}\right]\Bigg\rangle_J
    \end{aligned}
\end{align}
Now we perform the Gaussian integral over the measure to get the average, defining $\Vec{x}_{i,j}=\begin{pmatrix}
    \frac{y_i^{\alpha*}z_j^{\beta}+z_j^{\beta*}y_i^{\alpha}}{2} & \frac{y_j^{\alpha*}z_i^{\beta}+z_i^{\beta*}y_j^{\alpha}}{2}
\end{pmatrix}^T$, we have that the integral for the average starts looking like 
\begin{align}
    \begin{aligned}
        \int\prod_{i\alpha}\left(\frac{d^2z_i^\alpha d^2y_i^\alpha}{2\pi^2}\right)\prod_{ij}d\Vec{R}_{i,j}&\sqrt{\frac{NM}{(2\pi)^{NM}\Delta}}\exp\left[-\sum_{i\alpha}(y_i^{\alpha*}y_i^\alpha+\epsilon z_i^{\alpha*}z_i^\alpha+i\omega^* z_i^{\alpha*}y_i^\alpha+i\omega y_i^{\alpha*}z_i^\alpha)\right]\\&\times\exp\left[-\frac{NM}{4}\sum_{i,j}(\Vec{R}_{i,j}^T-\Vec{\mu}^T){\bf{\Gamma}}^{-1}(\Vec{R}_{i,j}-\Vec{\mu})+i\sum_{i,j}\Vec{R}_{i,j}^T\Vec{x}_{i,j}\right]\\
        =\int\prod_{i\alpha}\left(\frac{d^2z_i^\alpha d^2y_i^\alpha}{2\pi^2}\right)\prod_{ij}d\Vec{R}_{i,j}&\sqrt{\frac{NM}{(2\pi)^{NM}\Delta}}\exp\left[-\sum_{i\alpha}(y_i^{\alpha*}y_i^\alpha+\epsilon z_i^{\alpha*}z_i^\alpha+i\omega^* z_i^{\alpha*}y_i^\alpha+i\omega y_i^{\alpha*}z_i^\alpha)\right]\\&\times\exp\left[-\frac{NM}{4}\sum_{i,j}
        \Vec{R}_{i,j}^T{\bf{\Gamma}}^{-1}\Vec{R}_{i,j}+\sum_{i,j}\Vec{R}_{i,j}^T(i\Vec{x}_{i,j}+\frac{NM}{2}{\bf{\Gamma}}^{-1}\Vec{\mu})-\frac{NM}{4}\sum_{i,j}\Vec{\mu}^T{\bf{\Gamma}}^{-1}\Vec{\mu}\right]\\
        =\int\prod_{i\alpha}\left(\frac{d^2z_i^\alpha d^2y_i^\alpha}{2\pi^2}\right)\exp&\left[-\sum_{i\alpha}(y_i^{\alpha*}y_i^\alpha+\epsilon z_i^{\alpha*}z_i^\alpha+i\omega^* z_i^{\alpha*}y_i^\alpha+i\omega y_i^{\alpha*}z_i^\alpha)\right]\\&\times\exp\left[-\frac{1}{NM}\sum_{i,j}\Vec{x}_{i,j}^T{\bf{\Gamma}}\Vec{x}_{i,j}+i\sum_{i,j}\Vec{\mu}^T\Vec{x}_{i,j}\right]\\
        =\int\prod_{i\alpha}\left(\frac{d^2z_i^\alpha d^2y_i^\alpha}{2\pi^2}\right)\exp&\left[-\sum_{i\alpha}(y_i^{\alpha*}y_i^\alpha+\epsilon z_i^{\alpha*}z_i^\alpha+i\omega^* z_i^{\alpha*}y_i^\alpha+i\omega y_i^{\alpha*}z_i^\alpha)+i\sum_{ij\alpha\beta}(\mu_{\alpha\beta}y_i^{\alpha*}z_j^\beta+\mu_{\alpha\beta}z_j^{\beta*}y_i^\alpha)\right]\\&\times\exp\left[-\frac{1}{NM}\sum_{i,j}\Vec{x}_{i,j}^T{\bf{\Gamma}}\Vec{x}_{i,j}\right]
    \end{aligned}
\end{align}
which is just a Gaussian integral, in $2M^2$ dimensions, performed for $N^2$ different vectors. Thus,
\begin{align}
    \begin{aligned}
       \exp(-MN\Phi(\omega))=\int\prod_{i\alpha}\left(\frac{d^2z_i^\alpha d^2y_i^\alpha}{2\pi^2}\right)&\exp\left[-\sum_{i\alpha}(y_i^{\alpha*}y_i^{\alpha}+\epsilon z_i^{\alpha*}z_i^{\alpha})\right]\\\times\exp\Bigg[-i\sum_{ij\alpha\beta}(z_j^{\beta*}(\omega^*\delta_{\alpha\beta}\delta_{ij}&-\mu_{\alpha\beta})y_i^{\alpha}+y_i^{\alpha*}(\omega\delta_{\alpha\beta}\delta_{ij}-\mu_{\alpha\beta})z_j^{\beta})\Bigg]\\
\times \exp\Bigg[-\frac{1}{4NM}\sum_{ij\alpha\beta\alpha'\beta'}(y_{i}^{\alpha}z_j^{\beta*}&\Psi^{\alpha'\beta'}_{\alpha\beta}y_i^{\alpha'*}z_j^{\beta'}+y_{i}^{\alpha*}z_j^{\beta}\Psi^{\alpha'\beta'}_{\alpha\beta}z_j^{\beta'*}y_i^{\alpha'}+y_{j}^{\alpha*}z_i^{\beta}\Psi^{\alpha'\beta'}_{\alpha\beta}z_i^{\beta'*}y_j^{\alpha'}+z_{i}^{\beta*}y_j^{\alpha}\Psi_{\alpha\beta}^{\alpha'\beta'}z_i^{\beta'}y_j^{\alpha'*})\Bigg] \\
\times \exp\Bigg[-\frac{1}{4NM}\sum_{ij\alpha\beta\alpha'\beta'}(y_{j}^{\alpha*}z_i^{\beta}&\Upsilon^{\alpha'\beta'}_{\alpha\beta}y_i^{\alpha'*}z_j^{\beta'}+y_{j}^{\alpha}z_i^{\beta*}\Upsilon^{\alpha'\beta'}_{\alpha\beta}z_j^{\beta'*}y_i^{\alpha'}+z_{j}^{\beta}y_i^{\alpha*}\Upsilon^{\alpha'\beta'}_{\alpha\beta}y_j^{\alpha'*}z_i^{\beta'}+z_{j}^{\beta*}y_i^{\alpha}\Upsilon_{\alpha\beta}^{\alpha'\beta'}z_i^{\beta'*}y_j^{\alpha'})\Bigg] 
    \end{aligned}\label{eq:postgausint}
\end{align} where $\Psi=\overline{\sigma}^2$ and $\Upsilon=\overline{\gamma}^2$
Now we define the order parameters
\begin{align}
    \begin{aligned}
u_{\alpha\beta}&=\frac{1}{N}\sum_{i}z_i^{\alpha*}z_i^{\beta},\qquad             &v_{\alpha\beta}=\frac{1}{N}\sum_{i}y_i^{\alpha*}y_i^{\beta},\\                w_{\alpha\beta}&=\frac{1}{N}\sum_{i}y_i^{\alpha*}z_i^\beta,\qquad &w_{\alpha\beta}^*=\frac{1}{N}\sum_{i}y_i^{\alpha}z_i^{\beta*}    \end{aligned}\label{eq:ordparams}
\end{align}
We can impose these definitions using the usual integral representation of the delta function
\begin{align}
    \begin{aligned}
          \delta(v_{\alpha\beta}-\frac{1}{N}\sum_{i}y_i^{\alpha*}y_i^\beta)&\propto\int d\hat{v}^*_{\beta\alpha}\exp\left[i\hat{v}^*_{\beta\alpha}(Nv_{\alpha\beta}-\sum_{i}y_i^{\alpha*}y_i^\beta)\right]\\
          \delta(u_{\alpha\beta}-\frac{1}{N}\sum_{i}z_i^{\alpha*}z_i^\beta)&\propto\int d\hat{u}^*_{\beta\alpha}\exp\left[i\hat{u}^*_{\beta\alpha}(Nu_{\alpha\beta}-\sum_{i}z_i^{\alpha*}z_i^\beta)\right]\\
          \delta(w_{\alpha\beta}-\frac{1}{N}\sum_{i}y_i^{\alpha*}z_i^\beta)&\propto\int d\hat{w}_{\beta\alpha}^*\exp\left[i\hat{w}^*_{\beta\alpha}(Nw_{\alpha\beta}-\sum_{i}y_i^{\alpha*}z_i^\beta)\right]
          \\
                \delta(w_{\alpha\beta}^*-\frac{1}{N}\sum_{i}y_i^{\alpha}z_i^{\beta *})&\propto\int d\hat{w}_{\alpha\beta}\exp\left[i\hat{w}_{\alpha\beta}(Nw^*_{\beta\alpha}-\sum_{i}y_i^{\beta}z_i^{\alpha*})\right]
    \end{aligned}\label{eq:deltafuncs}
\end{align}
 Inserting \cref{eq:deltafuncs,eq:ordparams} into \cref{eq:postgausint}, we are able to write
\begin{align}
    \exp(-NM\Phi(\omega))=\int\mathcal{D}\{\dots\}\exp\{NM(\Xi+\Theta+\Omega)\}
\end{align}
where $\mathcal{D}\{\dots\}$ is the integral over all order parameters and their hatted analogues, and
\begin{align}
    \begin{aligned}
        \Xi&=i\frac{1}{M}\sum_{\alpha\beta}(u_{\alpha\beta}\hat{u}^*_{\beta\alpha}+v_{\alpha\beta}\hat{v}^*_{\beta\alpha}+\hat{w}^*_{\beta\alpha}w_{\alpha\beta}+\hat{w}_{\alpha \beta}w^*_{\beta\alpha})\\
        \Theta&=-\frac{1}{M}\sum_{\alpha}(v_{\alpha\alpha}+\epsilon u_{\alpha\alpha}+i\omega w_{\alpha\alpha}+i\omega^* w_{\alpha\alpha}^*)+i\frac{1}{M}\sum_{\alpha\beta}\mu_{\alpha\beta}(w_{\alpha\beta}+w_{\alpha\beta}^*)\\&\;\;\;\;\;\;\;\;\;\;\;\;\;\;\;\;\;\;\;\;\;\;\;\;\;\;\;\;\;\;\;\;\;\;-\frac{1}{M^2}\sum_{\alpha\beta\alpha'\beta'}\left(v_{\alpha'\alpha}u_{\beta\beta'}\Psi_{\alpha\beta}^{\alpha'\beta'}+\frac{1}{2}(w_{\alpha'\beta}w_{\alpha\beta'}+w_{\alpha\beta'}^*w^*_{\alpha'\beta})\Upsilon_{\alpha\beta}^{\alpha'\beta'}\right)\\
        \Omega&=\frac{1}{M}\ln{\int\left(\prod_{\alpha}\frac{d^2z^\alpha d^2y^\alpha}{2\pi^2}\right)\exp\{-i\sum_{\alpha\beta}(\hat{u}^*_{\beta\alpha}z^{\beta}z^{\alpha*}+\hat{v}^*_{\beta\alpha}y^{\beta}y^{\alpha*}+\hat{w}_{\beta\alpha}^*y^{\alpha*}z^{\beta}+\hat{w}_{\alpha\beta}y^{\beta}z^{\alpha*})\}}
    \end{aligned}
\end{align}
Perhaps we can rewrite the last one as
\begin{align}
    \begin{aligned}
        \Omega&=\frac{1}{M}\ln\int\left(\prod_{\alpha}\frac{d^2\Vec{z}^\alpha d^2\Vec{y}^\alpha}{2\pi^2}\right)\exp\Bigg[-i\begin{pmatrix}
            \Vec{y}^* & \Vec{z}^*
        \end{pmatrix}\begin{pmatrix}
            \hat{v}^*_{\beta\alpha}& \hat{w}_{\beta\alpha}^* \\ \hat{w}_{\alpha\beta} & \hat{u}^*_{\beta\alpha}
        \end{pmatrix}\begin{pmatrix}
            \Vec{y} \\ \Vec{z}
        \end{pmatrix}\Bigg]\\
            &=\frac{1}{M}\ln\int\left(\frac{d^2\Vec{V}}{(2\pi^2)^M}\right)\exp\left[-i\Vec{V}^{T*}\mathbb{A}\Vec{V}\right]\\
            &=-\frac{1}{M}\ln{\rm det}i\mathbb{A}\\&=-\frac{1}{M}\ln{\rm det}(\hat{w}^\dag\hat{w}-\hat{v}^\dag\hat{u}^\dag)
    \end{aligned}
\end{align}
where we have combined the integration variables $z^\alpha y^\alpha$ into one long vector $\Vec{V}$ of size $2M$ and performed a complex Gaussian integration, with
\begin{align}
    \begin{aligned}
        \mathbb{A}&=\begin{pmatrix}
            \hat{v}^*_{\beta\alpha} & \hat{w}_{\beta\alpha}^* \\ \hat{w}_{\alpha\beta} & \hat{u}^*_{\beta\alpha}
            \end{pmatrix}
    \end{aligned}
\end{align}
where $\hat{w},\hat{v}$, and $\hat{u}$ are $M\times M$ complex matrices. Now we need to extremise the potentials with respect to the hatted variables. 
\begin{align}
    \begin{aligned}
        \frac{\partial(\Xi+\Theta+\Omega)}{\partial\hat{u}^*_{\beta\alpha}}&=iu_{\alpha\beta}-{\rm Tr}\left[\mathbb{A}^{-1}\cdot\begin{pmatrix}
            0 & 0 \\0 & 1_{\alpha\beta}
        \end{pmatrix}\right]=0\\
        \frac{\partial(\Xi+\Theta+\Omega)}{\partial\hat{v}^*_{\beta\alpha}}&=iv_{\alpha\beta}-{\rm Tr}\left[\mathbb{A}^{-1}\cdot\begin{pmatrix}
           1_{\alpha\beta}& 0\\0&0
        \end{pmatrix}\right]=0\\
        \frac{\partial(\Xi+\Theta+\Omega)}{\partial\hat{w}_{\alpha\beta}}&=iw_{\beta\alpha}^*-{\rm Tr}\left[\mathbb{A}^{-1}\cdot\begin{pmatrix}
           0 & 0 \\ 1_{\alpha\beta} & 0
        \end{pmatrix}\right]=0\\
        \frac{\partial(\Xi+\Theta+\Omega)}{\partial\hat{w}^*_{\beta\alpha}}&=iw_{\alpha\beta}-{\rm Tr}\left[\mathbb{A}^{-1}\cdot\begin{pmatrix}
           0 & 1_{\alpha\beta} \\ 0& 0
        \end{pmatrix}\right]=0
    \end{aligned}
\end{align}
Letting $\mathbb{A}^{-1}=\mathcal{C}$, we see that
\begin{align}
    \begin{aligned}
        iu_{\alpha\beta}=\mathcal{C}^{22}_{\beta\alpha},\;\;
        iv_{\alpha\beta}=\mathcal{C}^{11}_{\beta\alpha},\;\;
        iw^*_{\beta\alpha}=\mathcal{C}^{12}_{\beta\alpha},\;\;
        iw_{\alpha\beta}=\mathcal{C}^{21}_{\beta\alpha}
        \implies \mathbb{A}^{-1}&=\begin{pmatrix}
            iv^T & iw^*\\iw^T&iu^T
        \end{pmatrix}
    \end{aligned}
\end{align}
This might add some helpful expressions to our repertoire if we remark that
\begin{align}
    \begin{aligned}
        \mathbb{A}\mathbb{A}^{-1}&=\begin{pmatrix}
            i\hat{v}^\dag v^T+i\hat{w}^\dag w^T&i\hat{v}^\dag w^*+i\hat{w}^\dag u^T\\i\hat{w}v^T+i\hat{u}^\dag w^T&i\hat{w}w^*+i\hat{u}^\dag u^T
        \end{pmatrix}=\mathbb{1}_{2M}
    \end{aligned}\label{eq:AAinv=1}
\end{align}
Let's carry on to the extremisation of the potentials with respect to the order parameters.
\begin{align}
    \begin{aligned}
        \frac{\partial (\Xi+\Theta+\Omega)}{\partial u_{\alpha\beta}}&=i\hat{u}_{\beta\alpha}^*-\frac{1}{M}\sum_{\alpha'\beta'}\Psi_{\beta'\beta}^{\alpha'\alpha}v_{\beta'\alpha'}-\delta_{\alpha\beta}\epsilon=0\\            \frac{\partial (\Xi+\Theta+\Omega)}{\partial v_{\alpha\beta}}&=i\hat{v}_{\beta\alpha}^*-\frac{1}{M}\sum_{\alpha'\beta'}\Psi_{\beta\beta'}^{\alpha\alpha'}u_{\beta'\alpha'}-\delta_{\alpha\beta}=0\\
        \frac{\partial (\Xi+\Theta+\Omega)}{\partial w_{\alpha\beta}}&=i\hat{w}_{\beta\alpha}^*-\frac{1}{M}\sum_{\alpha'\beta'}w_{\alpha'\beta'}\Upsilon^{\alpha'\beta}_{\alpha\beta'}-i(\omega\delta_{\alpha\beta}-\mu_{\alpha\beta})=0\\
        \frac{\partial (\Xi+\Theta+\Omega)}{\partial w^*_{\beta\alpha}}&=i\hat{w}_{\alpha\beta}-\frac{1}{M}\sum_{\alpha'\beta'}w_{\beta'\alpha'}^*\Upsilon^{\beta\alpha'}_{\beta'\alpha}-i(\omega^*\delta_{\alpha\beta}-\mu_{\beta\alpha})=0
    \end{aligned}\label{eq:hatremovals}
\end{align}
Rewriting these in a different notation,
\begin{align}
    \begin{aligned}
        i\hat{u}^\dag&=\frac{1}{M}\Psi \cdot v+\mathbb{1}\epsilon\\
        i\hat{v}^\dag&=\frac{1}{M}u\cdot\Psi +\mathbb{1}\\
        i\hat{w}^\dag&=\frac{1}{M}w^T\cdot\Upsilon^T+i(\mathbb{1}\omega-\mu)\\
        i\hat{w}&=\frac{1}{M}\Upsilon^T\cdot w^{*}+i(\mathbb{1}\omega^*-\mu^T)
    \end{aligned}
\end{align}where the dot signifies this higher dimensional matrix product, and the transpose on $\Upsilon$ only switches its upper indices. Inserting these relations into \cref{eq:AAinv=1}, we have
\begin{align}
\begin{aligned}
    \sum_{\alpha'\beta' \beta}(\Psi^{\alpha\alpha'}_{\beta\beta'}u_{\beta'\alpha'}v_{\gamma\beta}+w_{\gamma\beta}w_{\alpha'\beta'}\Upsilon^{\alpha'\beta}_{\alpha\beta'})+ Mv_{\gamma\alpha}+iM\sum_{\beta}w_{\gamma\beta}(\omega\delta_{\alpha\beta}-\mu_{\alpha\beta})&=M\delta_{\alpha\gamma}\\
    \sum_{\alpha'\beta'\beta}(\Psi^{\alpha\alpha'}_{\beta\beta'}u_{\beta'\alpha'}w^*_{\beta\gamma}+u_{\gamma\beta}w_{\alpha'\beta'}\Upsilon^{\alpha'\beta}_{\alpha\beta'})+ Mw^*_{\alpha\gamma}+iM\sum_{\beta}u_{\gamma\beta}(\omega\delta_{\alpha\beta}-\mu_{\alpha\beta})&=0\\
    \sum_{\alpha'\beta' \beta}(v_{\gamma\beta}w^*_{\beta'\alpha'}\Upsilon^{\beta\alpha'}_{\beta'\alpha}+w_{\gamma\beta}\Psi^{\alpha' \alpha}_{\beta'\beta}v_{\beta'\alpha'})+iM\sum_{\beta}v_{\gamma\beta}(\omega^*\delta_{\alpha\beta}-\mu_{\beta\alpha} )+M\epsilon w_{\gamma\alpha}&=0\\
    \sum_{\alpha'\beta'\beta}(w^*_{\beta\gamma}w^*_{\beta'\alpha'}\Upsilon^{\beta\alpha'}_{\beta'\alpha}+u_{\gamma\beta}\Psi^{\alpha'\alpha}_{\beta'\beta}v_{\beta'\alpha'})+iM\sum_{\beta}w^*_{\beta\gamma}(\omega^*\delta_{\alpha\beta}-\mu_{\beta\alpha} )+M\epsilon u_{\gamma\alpha}&=M\delta_{\alpha\gamma}
\end{aligned}\label{eq:4eqssumform}
\end{align}
which can be rewritten in the aforementioned matrix notation for intuitive understanding as
\begin{align}
    \begin{aligned}
        (u\cdot\Psi)v^T+Mv^T+(w^T\cdot\Upsilon^T)w^T+iM(\omega\mathbb{1}-\mu)w^T&=M\mathbb{1}\\
        (u\cdot\Psi)w^*+Mw^*+(w^T\cdot\Upsilon^T)u^T+iM(\omega\mathbb{1}-\mu) u^T&=0\\
        (\Upsilon^T\cdot w^{*})v^T+iM(\omega^*\mathbb{1}-\mu^T)v^T+(\Psi\cdot v)w^T+Mw^T\epsilon&=0\\
        (\Upsilon^T\cdot w^{*})w^*+iM(\omega^*\mathbb{1}-\mu^T)w^*+(\Psi\cdot v)u^T+Mu^T\epsilon&=M\mathbb{1}
    \end{aligned}\label{eq:4eqsmatform}
\end{align}
Now the resolvent is the derivative of the potential, which at the saddle point reads
\begin{align}
    G(\omega,\omega^*)=\frac{\partial\Phi(\omega,\omega^*)}{\partial\omega} 
    =i\frac{1}{M}{\rm Tr}\left[w(\omega,\omega^*)\right]\label{eq:resolvent}, 
\end{align}
and similarly
\begin{align}
    G^*(\omega,\omega^*)=\frac{\partial\Phi(\omega,\omega^*)}{\partial\omega^*} 
    =i\frac{1}{M}{\rm Tr}\left[w^*(\omega,\omega^*)\right]\label{eq:resolvent2}. 
\end{align}
Notice that (a bit strangely) $i w = (i w^*)^*$ as emphasized in \cite{baron2020dispersal}.

The spectral density is then the derivative of the resolvent
\begin{align}
    2\pi\rho={\rm Re}\left[\frac{\partial G}{\partial \omega^*}\right]={\rm Re}\left[\frac{\partial G^*}{\partial \omega}\right].
\end{align}
Hence if $w$ is only a function of $\omega$ (and $w^*$ is only a function of $\omega^*$), we get that the derivative of $G$ with respect to $\omega^*$ is equal to zero, defining regions with a zero density of eigenvalues 
\subsection{Conditions for boundary of the bulk of the spectrum}

Let us first study the third line of Eq.
\eqref{eq:4eqssumform}. When $\epsilon \to 0$, it is a linear equation for $v$ that can be written
\begin{equation} \label{L_equation}
    \sum_{\mu\nu} K_{\alpha\gamma}^{\mu \nu} v_{\mu \nu} = -i\omega^*v_{\alpha\gamma}
\end{equation}
with a linear operator $K$ defined as
\begin{equation}
K_{\alpha\gamma}^{\mu\nu} = \frac{1}{M}\sum_{\alpha'\beta'}\Upsilon^{\nu\alpha'}_{\beta'\gamma}w_{\beta'\alpha'}^*\delta_{\alpha\mu}+\frac{1}{M}\sum_\beta w_{\alpha\beta}\Psi^{\nu \gamma}_{\mu\beta}-i\delta_{\mu\alpha}\mu_{\nu\gamma}.
\end{equation}
Eq. \eqref{L_equation} always has the trivial solution $v \equiv 0$. If this is the case, one immediately observes that the first and last lines of Eq. \eqref{L_equation} imply that $G$ (resp. $G^*$) is an analytic function of $\omega$ (resp. $\omega^*$). Hence the density of eigenvalues is certainly zero when $v=0$. 

Another possibility is that $v$ is a non-trivial eigenvector of $K$ with eigenvalue $-i \omega^*$. We assume, as in \cite{baron2020dispersal}, that $v$ and $w^*$ are continuous at the boundary limiting the complex plane domain where the density is non zero. (Note that this does not imply that the density itself is continuous -- and in fact, it is not in generic cases.) One can then determine the matrices $w^*$ appearing in $K$ using the second line of 
\begin{align}
\begin{aligned}
    \sum_{\alpha'\beta' \beta}w_{\gamma\beta}w_{\alpha'\beta'}\Upsilon^{\alpha'\beta}_{\alpha\beta'}+iM\sum_{\beta}(\omega\delta_{\beta\alpha}-\mu_{\alpha\beta} )w_{\gamma\beta}&=M\delta_{\alpha\gamma}\\\sum_{\alpha'\beta'\beta}w^*_{\beta\gamma}w^*_{\beta'\alpha'}\Upsilon^{\beta\alpha'}_{\beta'\alpha}+iM\sum_{\beta}w^*_{\beta\gamma}(\omega^*\delta_{\alpha\beta}-\mu_{\beta\alpha} )&=M\delta_{\alpha\gamma}
\end{aligned}\label{eq:wweqs}
\end{align}
which can be simplified as 
\begin{align}
    \begin{aligned}
\sum_\beta w^*_{\beta\gamma}(\sum_{\alpha'\beta'}w^*_{\beta'\alpha'}\Upsilon^{\beta\alpha'}_{\beta'\alpha}+iM(\omega^*\delta_{\alpha\beta}-\mu_{\beta\alpha}))&=M\delta_{\alpha\gamma}\\
\implies\sum_{\gamma\beta}(w^*)^{-1}_{\gamma\eta}w^*_{\beta\gamma}\sum_{\alpha'\beta'}w^*_{\beta'\alpha'}\Upsilon^{\beta\alpha'}_{\beta'\alpha}+\sum_{\beta\gamma}i(w^*)^{-1}_{\gamma\eta}M(\omega^*\delta_{\alpha\beta}-\mu_{\beta\alpha})w^*_{\beta\gamma}&=M\sum_{\gamma}\delta_{\alpha\gamma}(w^*)^{-1}_{\gamma\eta}\\
\implies\sum_{\alpha'\beta'}w^*_{\beta'\alpha'}\Upsilon^{\beta\alpha'}_{\beta'\alpha}&=M(w^*)^{-1}_{\alpha\beta}-iM(\omega^*\delta_{\alpha\beta}-\mu_{\beta\alpha})
    \end{aligned}
\end{align}
This expression can be inserted in the eigenvalue equation for $v$:
\begin{align}
    \begin{aligned}
        \sum_{\mu\nu}(K_{\alpha\gamma}^{\mu\nu}+i\omega^*\delta_{\alpha\mu}\delta_{\gamma\nu})v_{\mu\nu}&=0\\
        \implies\sum_{\mu\nu}\left[ \frac{1}{M}\sum_{\alpha'\beta'}\Upsilon_{\beta'\gamma}^{\nu\alpha'}w_{\beta'\alpha'}^*\delta_{\alpha\mu}+\frac{1}{M}\sum_\beta w_{\alpha\beta}\Psi^{\nu \gamma}_{\mu\beta}+i(\omega^*\delta_{\alpha\mu}\delta_{\gamma\nu}-\mu_{\nu\gamma}\delta_{\alpha\mu})\right]v_{\mu\nu}&=0\\
        \implies\sum_{\mu\nu}\left[(w^*)^{-1}_{\gamma\nu}\delta_{\alpha\mu}-i(\omega^*\delta_{\gamma\nu}-\mu_{\nu\gamma})\delta_{\alpha\mu}+\frac{1}{M}\sum_\beta w_{\alpha\beta}\Psi^{\nu \gamma}_{\mu\beta}+i(\omega^*\delta_{\alpha\mu}\delta_{\gamma\nu}-\mu_{\nu\gamma}\delta_{\alpha\mu})\right]v_{\mu\nu}&=0\\
        \implies \sum_{\mu\nu}\left[\sum_{\gamma}w^*_{\eta\gamma}\left((w^*)^{-1}_{\gamma\nu}\delta_{\alpha\mu}+\frac{1}{M}\sum_\beta w_{\alpha\beta}\Psi^{\nu \gamma}_{\mu\beta}\right)\right]v_{\mu\nu}&=0\\
        \implies\Bigg\vert \delta_{\eta\nu}\delta_{\alpha\mu}+\frac{1}{M}\sum_{\beta\gamma} w_{\alpha\beta}\Psi^{\nu \gamma}_{\mu\beta}w^*_{\eta\gamma}\Bigg\vert&=0\\
    \end{aligned}
\end{align}
We now have the conditions defining the boundary as:
\begin{align}
    \begin{aligned}
        \Bigg\vert \delta_{\eta\nu}\delta_{\alpha\mu}+\frac{1}{M}\sum_{\beta\gamma} w_{\alpha\beta}\Psi^{\nu \gamma}_{\mu\beta}w^*_{\eta\gamma}\Bigg\vert&=0\\
         \sum_{\alpha'\beta' \beta}w_{\gamma\beta}w_{\alpha'\beta'}\Upsilon^{\alpha'\beta}_{\alpha\beta'}+iM\sum_{\beta}(\omega\delta_{\beta\alpha}-\mu_{\alpha\beta} )w_{\gamma\beta}&=M\delta_{\alpha\gamma}\\
        iw_{\alpha\beta}&=(iw_{\alpha\beta}^*)^*
    \end{aligned}\label{eq:allcondssupp}
\end{align}
Note that the determinant needs to be calculated by converting the object in 4 indices to 2 indices, grouping $\alpha\gamma$ and $\mu\nu$ separately. We are also able to confirm that all matrices $v$ that are ``eigenvectors'' of the object $\Psi w^*w$ will also be hermitian. We would get the same equations whenever 
\begin{align}
    u^T(\Psi\cdot v)=\sum_{\alpha'\beta'\beta}u_{\beta\alpha}\Psi^{\alpha'\beta}_{\beta'\gamma}v_{\beta'\alpha'}=0\label{eq:zerocond}
\end{align}
since in this case the term $u^T(\Psi \cdot v)$ drops out of equation on $w^*$, even when $v \neq 0$ but belongs to the null space of $\Psi$ (the case $u=0$ leads to inconsistencies). 
\begin{figure}
    \centering
    \includegraphics[width=\linewidth]{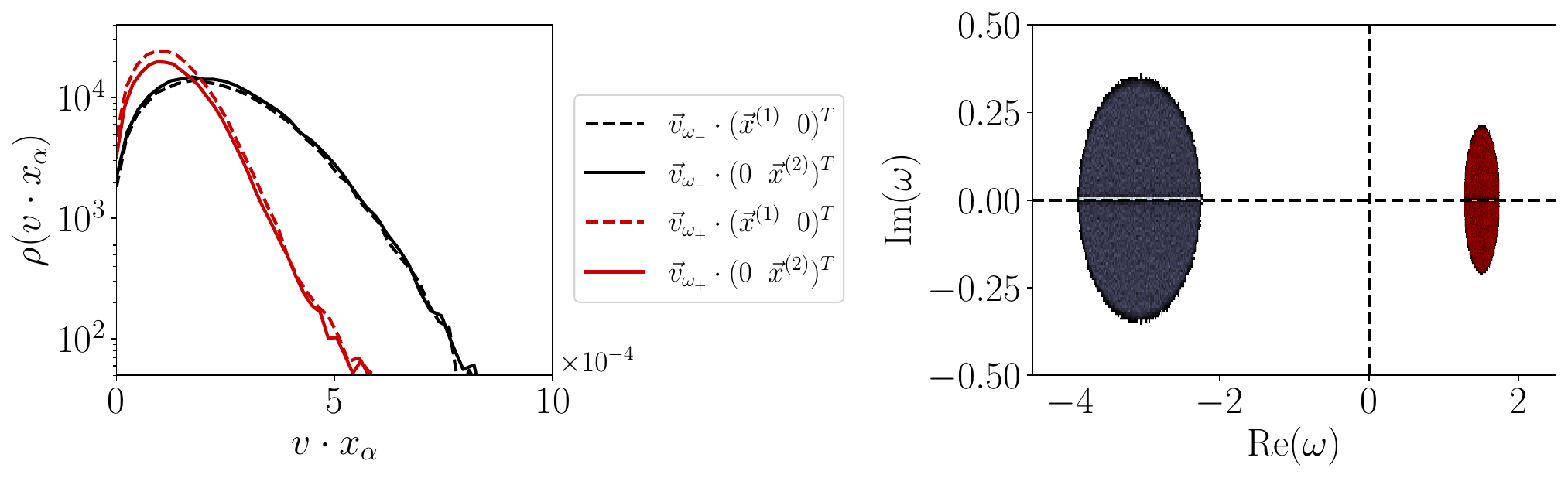}
    \caption{Here we see the regional properties of the eigenvectors for the Lotka-Volterra model on two islands. On the left, the red lines are the distributions of the dot products of the eigenvectors corresponding to the positive eigenvalues (red region on right) with vectors corresponding to the two variables $\Vec{x}^{(1)}$ and $\Vec{x}^{(2)}$, and the black lines are the same for the eigenvectors corresponding to the negative eigenvalues (grey region on the right). We see that one set of vectors is more widely distributed among the two spaces, while the other is somewhat localised in comparison. We see this behaviour in cases where there are two eigenvalue regions separated by boundaries given by our analytical equations, even if they overlap entirely, as in \cref{fig:spectra_difftauandgam}.}
    \label{fig:eigenvectorlocalization}
\end{figure}
Hence when we solve numerically eqs. \eqref{eq:allcondssupp} we expect to find lines that do not correspond to the boundary delimiting the spectrum of our initial problem. As discussed in the main text, we conjecture that these lines correspond to changes in the structure of the eigenvectors of the stability matrix. We illustrate this in \cref{fig:eigenvectorlocalization}, where the components of one set of eigenvectors is seen to be widely spread among the spaces of both variables in the system ($x^{(1)}$ and $x^{(2)}$), while the other is localised in a seemingly statistically similar manner in both.

\section{The Random Rotation Trick\label{sec:rotappendix}}

Let us consider the following randomly rotated matrix:
\begin{align}
    \begin{aligned}
        R_{ab}=\sum_{ij}W_{ai}W_{bj}(\delta_{ij}\sum_mf_{im}+f_{ij})
    \end{aligned}
\end{align}
which gives that the correlations between various terms of the rotated matrix then are
\begin{align}
    \begin{aligned}
        \overline{R_{ab}R_{cd}}&=\sum_{ijkl}W_{ai}W_{bj}W_{ck}W_{dl}\overline{(\delta_{ij}\sum_mf_{im}+f_{ij})(\delta_{kl}\sum_nf_{kn}+f_{kl})}\\
        &=\sum_{ijkl}W_{ai}W_{bj}W_{ck}W_{dl}(\delta_{ij}\delta_{kl}\sum_{mn}\overline{f_{im}f_{kn}}+\delta_{ij}\sum_m\overline{f_{im}f_{kl}}+\delta_{kl}\sum_n\overline{f_{ij}f_{kn}}+\overline{f_{ij}f_{kl}})\\
        &=\sigma^2\sum_{ijkl}W_{ai}W_{bj}W_{ck}W_{dl}(\delta_{ij}\delta_{kl}(\delta_{ik}N+\tau)+\delta_{ij}(\delta_{ik}+\delta_{il}\tau)+\delta_{kl}(\delta_{ik}+\delta_{jk}\tau)+(\delta_{ik}\delta_{jl}+\delta_{il}\delta_{jk}\tau))\\
        &=\sigma^2\Bigg(N\sum_iW_{ai}W_{bi}W_{ci}W_{di}+\sum_iW_{ai}W_{bi}W_{ci}\sum_lW_{dl}+\sum_iW_{ai}W_{bi}W_{di}\sum_lW_{ck}+\sum_iW_{ai}W_{ci}W_{di}\sum_jW_{bj}\\&+\sum_{ik}W_{ai}W_{bi}W_{ck}W_{dk}\tau+\sum_iW_{ai}\sum_jW_{bj}W_{cj}W_{dj}+\sum_iW_{ai}W_{ci}\sum_jW_{bj}W_{dj}+\sum_iW_{ai}W_{di}\sum_jW_{bj}W_{cj}\tau\Bigg)
    \end{aligned}
\end{align}
If $N$ is large enough the vector overlaps are as follows
\begin{align}
    \begin{aligned}
       \sum_iW_{ai}&=\eta_a\\
       \sum_{i}W_{ai}W_{bi}&=\delta_{ab}\\
       \sum_{i}W_{ai}W_{bi}W_{ci}&=\frac{\eta_{3}}{N}\\
       \sum_iW_{ai}W_{bi}W_{ci}W_{di}&=\frac{1}{N}\delta_{ab}\delta_{cd}+\frac{1}{N}\delta_{ac}\delta_{bd}+\frac{1}{N}\delta_{ad}\delta_{bc}+\frac{\eta_4}{N^{3/2}}
    \end{aligned}
\end{align}
where the different $\eta$'s are random variables with $0$ mean and finite variance. These expressions can be combined to then give the result in the main text, i.e.
\begin{align}
    \begin{aligned}
        \overline{R_{ab}R_{cd}}&=2\sigma^2\delta_{ac}\delta_{bd}+\sigma^2(1+\tau)\delta_{ad}\delta_{bc}+\mathcal{O}(\frac{1}{N}).
    \end{aligned}
\end{align}
 
\section{Two Island Lotka-Volterra system\label{appsec:2LV}}
Consider the two island Lotka-Volterra equations
\begin{align}
    \begin{aligned}
        \Dot{x}_i&=\gamma x_i(\sum_jJ_{ij}(x_j+\tau_1y_j)-(x_i+\tau_2y_i)+\kappa_x)\\
        \Dot{y}_i&=\gamma y_i(\sum_jJ_{ij}(y_j+\tau_1x_j)-(y_i+\tau_2x_i)+\kappa_y)
    \end{aligned}
\end{align}
which can be expressed in matrix form as
\begin{align}
    \begin{pmatrix}
        \Dot{\delta x}\\\Dot{\delta y}
    \end{pmatrix}=L\begin{pmatrix}
        \delta x\\\delta y
    \end{pmatrix}, \qquad {\rm with} \qquad L=\begin{pmatrix}
        L_{xx}&L_{xy}\\L_{yx}&L_{yy}
    \end{pmatrix},
\end{align} where the terms of the stability matrix are given by
\begin{align}
    \begin{aligned}
        L^{ij}_{xx}&=\gamma\delta_{ij}(\sum_kJ_{ik}(x^*+\tau_1y^*)-2x^*-\tau_2y^*+\kappa_x)+\gamma x_i^* J_{ij}\\
        L^{ij}_{xy}&=-\delta_{ij}\gamma x_i^*\tau_2+\gamma x_i^*\tau_1 J_{ij}\\
        L_{yx}^{ij}&=-\delta_{ij}\gamma y_i^*\tau_2+\gamma y_i^*\tau_1 J_{ij}\\
        L_{yy}^{ij}&=\gamma\delta_{ij}(\sum_kJ_{ik}(y^*+\tau_1x^*)-2y^*-\tau_2x^*+\kappa_y)+\gamma y_i^* J_{ij}
    \end{aligned}
\end{align}
around steady state we'll also have $x_i^*\approx\frac{\kappa_x-\tau_2\kappa_y}{1-\tau_2^2}, y_i^*\approx \frac{\kappa_y-\tau_2\kappa_x}{1-\tau_2^2}$.
which lets us express all terms of our rotated matrix $R$ in terms of the system parameters,

\begin{align}
    \begin{aligned}R_{ij}^{11}&=\sum_{kl}w^{1,x}_{i,k}L^{xx}_{kl}w^{1,x}_{j,l}
    =-\delta_{ij}x^*\gamma+\gamma\sum_{k\mu}w_{ik}w_{jk}J_{k\mu}(x^*+\tau_1y^*)+\gamma\sum_{kl}w_{ik}w_{jl}x^*J_{kl}\\
    R_{ij}^{12}&=\sum_{kl}w^{1,x}_{i,k}L^{xy}_{kl}w^{2,y}_{j,l} =-\delta_{ij}x^*\tau_2\gamma+\gamma\sum_{kl}w_{ik}w_{jl}x^*\tau_1J_{kl}\\
    R_{ij}^{21}&=\sum_{kl}w^{2,y}_{i,k}L^{yx}_{kl}w^{1,x}_{j,l}-\delta_{ij}y^*\tau_2\gamma+\gamma\sum_{kl}w_{ik}w_{jl}y^*\tau_1J_{kl}\\
    R_{ij}^{22}&=\sum_{kl}w^{2,y}_{i,k}L^{yy}_{kl}w^{2,y}_{j,l}=-\delta_{ij}y^*\gamma+\gamma\sum_{k\mu}w_{ik}w_{jk}J_{k\mu}(y^*+\tau_1x^*)+\gamma\sum_{kl}w_{ik}w_{jl}y^*J_{kl}
    \end{aligned}
\end{align}
where the choice made on the vector basis $V=\delta\times W$ means $w^{1,y}=0$ and $w^{2,x}=0$, and the basis $w$ is orthogonal and normalised to $1$. So the statistics of the matrix are 
\begin{align}
    \begin{aligned}
        \overline{R}_{ij}&=-\delta_{ij}\gamma\begin{pmatrix}
            x^*&x^*\tau_2\\y^*\tau_2&y^*     \end{pmatrix}\\
            MN\overline{R_{ij}R_{kl}}&=NM\delta_{ik}\delta_{jl}\sigma_J^2\gamma^2\begin{pmatrix}
                (x^*+y^*\tau_1)^2+x^{*2} & \tau_1x^{*2}& \tau_1x^*y^*& (x^*+y^*\tau_1)(y^*+x^*\tau_1)+x^*y^*\\\tau_1x^{*2}& (x^*\tau_1)^2 & x^*y^*\tau_1^2& \tau_1x^*y^* \\
                \tau_1x^*y^* & x^*y^*\tau_1^2 & (y^*\tau_1)^2  &\tau_1y^{*2}\\
                (x^*+y^*\tau_1)(y^*+x^*\tau_1)+x^*y^*&\tau_1x^*y^*&\tau_1y^{*2}& (y^*+x^*\tau_1)^2+y^{*2} 
            \end{pmatrix}\\&+NM\delta_{il}\delta_{jk}\sigma_J^2\gamma^2\rho\begin{pmatrix}
                \frac{1}{\rho}(x^*+y^*\tau_1)^2+x^{*2} & \tau_1x^{*2}& \tau_1x^*y^*& \frac{1}{\rho}(x^*+y^*\tau_1)(y^*+x^*\tau_1)+x^*y^*\\\tau_1x^{*2}& (x^*\tau_1)^2 & x^*y^*\tau_1^2& \tau_1x^*y^* \\
                \tau_1x^*y^* & x^*y^*\tau_1^2 & (y^*\tau_1)^2  &\tau_1y^{*2}\\
                \frac{1}{\rho}(x^*+y^*\tau_1)(y^*+x^*\tau_1)+x^*y^*&\tau_1x^*y^*&\tau_1y^{*2}& \frac{1}{\rho}(y^*+x^*\tau_1)^2+y^{*2} 
            \end{pmatrix}
    \end{aligned}
\end{align}
So these are our $\Psi$ and  $\Upsilon$ matrices respectively.

\section{Good exchange economic model}
We modify the initial dynamical equations and rewrite them for exponents $\xi=\log p$ and $\eta=\log S$ 
\begin{align}
    \begin{aligned}
        \frac{d\xi_i}{dt}&=\frac{\kappa}{Ny}\left(\sum_j\Theta(\omega_{ji}-\xi_i-\eta_j)-Y\right)\\
        \frac{d\eta_i}{dt}&=\frac{\kappa}{Ny}\left(\sum_j\Theta(\omega_{ij}-\xi_j-\eta_i)e^{\xi_j-\xi_i}-\sum_j\Theta(\omega_{ji}-\xi_i-\eta_j)\right)
    \end{aligned}
\end{align}
where $\omega=\log J$. Looking for linear perturbations of the sort $\xi_i=\xi_i^*+\delta\xi_i$ and $\eta_i=\eta_i^*+\delta\eta_i$, we would have,
\begin{align}
    \begin{aligned}
        \frac{d\delta\xi_i}{dt}&=\frac{\kappa}{Ny}(\sum_j\Theta(\omega_{ji}-\xi^*_i-\eta^*_j-\delta\xi_i-\delta\eta_j)-Y)\\
        \implies \frac{d\delta\xi_i}{dt}&=-\frac{\kappa}{Ny}(\sum_j\delta(\omega_{ji}-\xi^*_i-\eta^*_j)(\delta\xi_i+\delta\eta_j))\\
        \implies \frac{d\delta\xi_i}{dt}&=-\delta\xi_i\frac{\kappa}{Ny}\sum_j\delta(\omega_{ji}-\xi^*_i-\eta^*_j)-\frac{\kappa}{Ny}\sum_j\delta(\omega_{ji}-\xi^*_i-\eta^*_j)\delta\eta_j
    \end{aligned}
\end{align}
Similarly,
\begin{align}
    \begin{aligned}
        \frac{d\delta\eta_i}{dt}=\frac{\kappa}{Ny}\sum_j\Theta&(\omega_{ij}-\xi^*_j-\eta^*_i-\delta\xi_j-\delta\eta_i)e^{\xi^*_j-\xi^*_i+\delta\xi_j-\delta\xi_i}-\frac{\kappa}{Ny}\sum_j\Theta(\omega_{ji}-\xi\*_i-\eta^*_j-\delta\xi_i-\delta\eta_j)\\
        \implies \frac{d\delta\eta_i}{dt}=\frac{\kappa}{Ny}\sum_j\Theta&(\omega_{ij}-\xi^*_j-\eta^*_i)e^{\xi^*_j-\xi^*_i}(\delta\xi_j-\delta\xi_i)-\frac{\kappa}{Ny}\sum_j\delta(\omega_{ij}-\xi^*_j-\eta^*_i)e^{\xi^*_j-\xi^*_i}(\delta\xi_j+\delta\eta_i)\\&+\frac{\kappa}{Ny}\sum_j\delta(\omega_{ji}-\xi\*_i-\eta^*_j)(\delta\xi_i+\delta\eta_j)\\
        \implies \frac{d\delta\eta_i}{dt}=\frac{\kappa}{Ny}\sum_j(\Theta&(\omega_{ij}-\xi^*_j-\eta^*_i)-\delta(\omega_{ij}-\xi^*_j-\eta^*_i))e^{\xi^*_j-\xi^*_i}\delta\xi_j\\
        &+\frac{\kappa}{Ny}\delta\xi_i\sum_j(\delta(\omega_{ji}-\xi^*_i-\eta^*_j)-\Theta(\omega_{ij}-\xi^*_j-\eta^*_i)e^{\xi^*_j-\xi^*_i})\\
        &+\frac{\kappa}{Ny}\sum_j\delta(\omega_{ji}-\xi\*_i-\eta^*_j)\delta\eta_j-\delta\eta_i\frac{\kappa}{Ny}\sum_j\delta(\omega_{ij}-\xi\*_j-\eta^*_i)e^{\xi_j^*-\xi_i^*}
    \end{aligned}
\end{align}
This gives the matrix equation for their perturbations
\begin{align}\Dot{
    \begin{pmatrix}
    \Vec{\delta\xi}\\\Vec{\delta\eta}
    \end{pmatrix}}=\begin{pmatrix}
    \mathbb{A} &\mathbb{B} \\\mathbb{C} & \mathbb{D}
    \end{pmatrix}\begin{pmatrix}
    \Vec{\delta\xi}\\\Vec{\delta\eta}
    \end{pmatrix}
\end{align}
where like before we have
\begin{align}
    \begin{aligned}
        \mathbb{A}_{ij}&=-\delta_{ij}\frac{\kappa}{Ny}\sum_k \theta'_{ki}\\
        \mathbb{B}_{ij}&=-\frac{\kappa}{Ny}\theta'_{ji}\\
        \mathbb{C}_{ij}&=-\delta_{ij}\frac{\kappa}{Ny}\sum_k\left[e^{\xi^*_k-\xi^*_i}\theta_{ik}-\theta'_{ki}\right]+\frac{\kappa}{Ny}e^{\xi^*_j-\xi^*_i}\left[\theta_{ij}-\theta'_{ij}\right]\\
        \mathbb{D}_{ij}&=-\delta_{ij}\frac{\kappa}{Ny}\sum_k \theta'_{ik}e^{\xi^*_k-\xi^*_i}+\frac{\kappa}{Ny}\theta'_{ji}
    \end{aligned}\label{eq:ABCDblockdefs}
\end{align}
where the function $\theta(\beta E_{ij})$ is the continuous form of the Heaviside theta function chosen, which in our case is again Fermi-Dirac. Taking average over disorder and expressing $\xi^*_i=\mu_\xi+\zeta_i/\sqrt{N}$ and $\eta^*_i=\mu_\eta+\epsilon_i/\sqrt{N}$, we can 
divide this into the constant and the fluctuating parts, the constant matrix given by
\begin{align}
\begin{aligned}
    \mathbb{A}'_{ij}&=-\delta_{ij}\frac{\kappa}{y}P_\omega^*\\
    \mathbb{B}'_{ij}&=-\frac{\kappa}{Ny}P_\omega^*\\
    \mathbb{C}'_{ij}&=-\delta_{ij}\frac{\kappa}{y}(P_>^*-P_\omega^*)+(P_>^*-P_\omega)\frac{\kappa}{Ny}\\
    \mathbb{D}'_{ij}&=-\delta_{ij}\frac{\kappa}{y}P_\omega^*+\frac{\kappa}{Ny}P_\omega^*
\end{aligned}
\end{align}
Rotating the matrix in the way prescribed in the main text with a double orthogonal basis $W_{(M)}\times W_{(N)}$, we will have
\begin{align}
\begin{aligned}
    \frac{N^2y^2}{\kappa^2}\overline{\mathbb{R}_{i^{\epsilon_1} j^{\epsilon_2}}\mathbb{R}_{i^{\epsilon_3} j^{\epsilon_4}}}=&
        \begin{pmatrix}
             \begin{pmatrix}
                 2 & 0 \\ -2 & 0
             \end{pmatrix} &
             \begin{pmatrix}
                 0 & 2 \\ 0 & -2
             \end{pmatrix} \\    
             \begin{pmatrix}
                 -2 & 0 \\ 10 & 0
             \end{pmatrix} &
             \begin{pmatrix}
                 0 & -2 \\ 0 & 10\end{pmatrix}\end{pmatrix}\frac{\beta P_\omega^*}{6} +\begin{pmatrix}
             \begin{pmatrix}
                 0 & 2 \\ 0 & 2
             \end{pmatrix} &
             \begin{pmatrix}
                 2 & 0 \\ -2 & 0
             \end{pmatrix} \\    
             \begin{pmatrix}
                 0 & -2 \\ -8 & 2
             \end{pmatrix} &
             \begin{pmatrix}
                 2 & 0 \\ 2 & -8\end{pmatrix}\end{pmatrix}P_\omega^2 \\&+\begin{pmatrix}
             \begin{pmatrix}
                 2 & 0 \\ 2 & 0
             \end{pmatrix} &
             \begin{pmatrix}
                 0 & -2 \\ 0 & -2
             \end{pmatrix} \\    
             \begin{pmatrix}
                 2 & 0 \\ 2 & 0
             \end{pmatrix} &
             \begin{pmatrix}
                 0 & -2 \\ 0 & -2\end{pmatrix}\end{pmatrix}(P_\omega/2-P_>P_\omega) +\begin{pmatrix}
             \begin{pmatrix}
                 1 & 1 \\ -1 & -1
             \end{pmatrix} &
             \begin{pmatrix}
                 1 & 1 \\ -1 & -1
             \end{pmatrix} \\    
             \begin{pmatrix}
                 -1 & -1 \\ 1 & 1
             \end{pmatrix} &
             \begin{pmatrix}
                 -1 & -1 \\ 1 & 1\end{pmatrix}\end{pmatrix}(2P_>P_\omega+P_>)  
\end{aligned}
\end{align} from which we can deduce $\Psi^{\epsilon_1\epsilon_2}_{\epsilon_3\epsilon_4}=NM\overline{\mathbb{R}_{i^{\epsilon_1}j^{\epsilon_2}}\mathbb{R}_{i^{\epsilon_3}j^{\epsilon_4}}}$, and
\begin{align}
\begin{aligned}
    \frac{N^2y^2}{\kappa^2}\overline{\mathbb{R}_{i^{\epsilon_1} j^{\epsilon_2}}\mathbb{R}_{j^{\epsilon_4} i^{\epsilon_3}}}=&
        \begin{pmatrix}
             \begin{pmatrix}
                 1 & -1 \\ 1 & -1
             \end{pmatrix} &
             \begin{pmatrix}
                 -1 & 1 \\ 3 & -3
             \end{pmatrix} \\    
             \begin{pmatrix}
                 1 & 3 \\ 1 & 3
             \end{pmatrix} &
             \begin{pmatrix}
                 -1 & -3 \\ 3 & 9\end{pmatrix}\end{pmatrix}(\frac{\beta P_\omega^*}{6}-P_\omega^2) +\begin{pmatrix}
             \begin{pmatrix}
                 -2 & 0 \\ 4 & -2
             \end{pmatrix} &
             \begin{pmatrix}
                 0 & 2 \\ 2 & -4
             \end{pmatrix} \\    
             \begin{pmatrix}
                 4 & 2 \\ -6 & 0
             \end{pmatrix} &
             \begin{pmatrix}
                 -2 & -4 \\ 0 & 6\end{pmatrix}\end{pmatrix}P_\omega P_> \\&+\begin{pmatrix}
             \begin{pmatrix}
                 3 & 0 \\ -2 & -1
             \end{pmatrix} &
             \begin{pmatrix}
                 0 & -3 \\ 1 & 2
             \end{pmatrix} \\    
             \begin{pmatrix}
                 -2 & 1 \\ -1 & 0
             \end{pmatrix} &
             \begin{pmatrix}
                 -1 & 2 \\ 0 & -1\end{pmatrix}\end{pmatrix}P_\omega +\begin{pmatrix}
             \begin{pmatrix}
                 1 & 1 \\ -1 & -1
             \end{pmatrix} &
             \begin{pmatrix}
                 1 & 1 \\-1 & -1
             \end{pmatrix} \\    
             \begin{pmatrix}
                 -1 & -1 \\ 1 & 1
             \end{pmatrix} &
             \begin{pmatrix}
                 -1 & -1 \\ 1 & 1\end{pmatrix}\end{pmatrix}(P_>-P_>^2)  
\end{aligned}
\end{align}
which is $\Upsilon^{\epsilon_1\epsilon_2}_{\epsilon_3\epsilon_4}=NM\overline{\mathbb{R}_{i^{\epsilon_1}j^{\epsilon_2}}\mathbb{R}_{i^{\epsilon_3}j^{\epsilon_4}}}$ multiplied by certain factors. For the spectra in \cref{fig:RRNmodelspectra}, we ignore the terms not containing $\beta$, as they are much smaller in comparison. 

\end{widetext}
\bibliography{apssamp}

\end{document}